\documentclass[conference]{IEEEtran}
\IEEEoverridecommandlockouts
\usepackage{cite}
\usepackage{amsmath,amssymb,amsfonts}
\usepackage[linesnumbered,ruled]{algorithm2e}
\usepackage{graphicx}
\usepackage{textcomp}
\usepackage{xcolor}
\usepackage{booktabs} 
\usepackage{multirow}
\usepackage{array}
\usepackage{pifont}
\usepackage[labelformat=simple]{subcaption}
\def\BibTeX{{\rm B\kern-.05em{\sc i\kern-.025em b}\kern-.08em
    T\kern-.1667em\lower.7ex\hbox{E}\kern-.125emX}}
\usepackage[hidelinks]{hyperref} 

\begin{document}

\title{A System Aware Resource Allocation for Distributed Workflows in Quantum Computing Environments
}

\author{
\IEEEauthorblockN{Abhishek Sawaika, Udaya Parampalli, Rajkumar Buyya}
\IEEEauthorblockA{\textit{
Quantum Cloud Computing and Distributed Systems (qCLOUDS) Lab},\\
\textit{School of Computing and Information System},\\
\textit{The University of Melbourne, Australia}\\
abhishek.sawaika@student.unimelb.edu.au, (udaya, rbuyya)@unimelb.edu.au}
}
\maketitle

\begin{abstract}

Rapid advancements in cloud based platforms providing access to quantum computing capabilities have opened up several challenges for efficient usage of these highly delicate and costly devices. Although most of the current systems use a priority based access protocol, they are unable to fully support reliable, efficient, and scalable execution of larger-scale applications. 
To overcome this limitation, we propose a comprehensive solution for efficient allocation of quantum programs to appropriate quantum devices, considering all the relevant cost metrics into account including, fidelity, execution time and communication overhead.
We also formulate use-cases for distributed quantum workflow and propose modified graph based algorithms to solve for allocation of such use-cases, assuming a hybrid classical-quantum network. Since hardware advancements in large standalone devices is an ongoing process, it is critical to investigate such distributed workflows to maximize the best utilization of current NISQ devices. Our empirical study shows that the proposed techniques perform better than state-of-the-art methods for almost all evaluation parameters, with average improvements of approximately $5\%$ in execution time, $30\%$ in communication overhead, $40\%$ in wait time and $2\%$ in fidelity, providing better solutions to efficient allocation strategies.

\end{abstract}


\section{Introduction}

\begin{figure}[h]
    \centering
    \includegraphics[width=\linewidth]{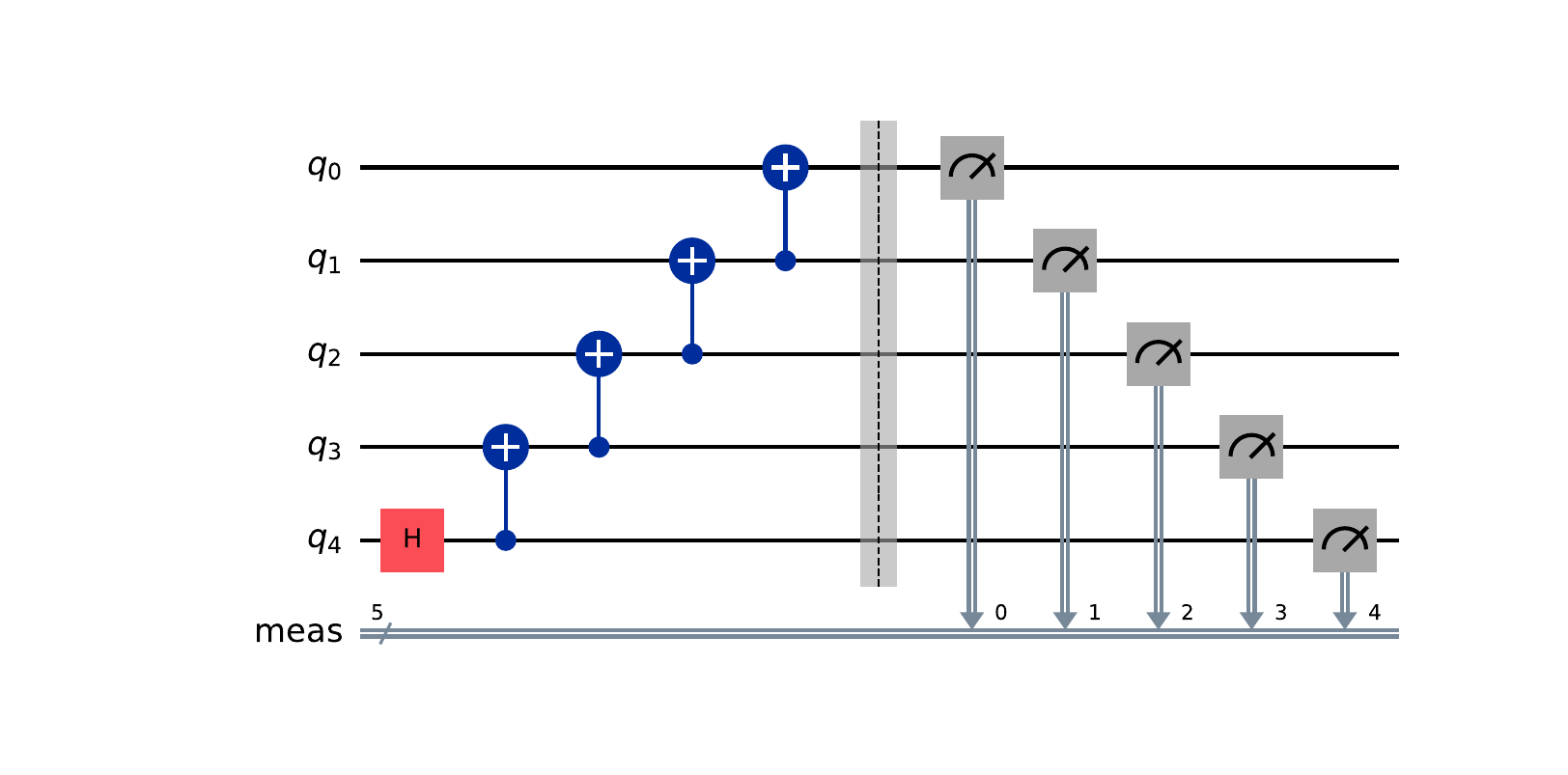}
    \caption{A sample quantum circuit (GHZ state preparation) with 5 qubits, depth=6, 1 single qubit operation (H), four 2-qubits operations (CNOT) and 5 measurements, representing a quantum program.}
    \label{fig:program}
\end{figure}

Quantum computing is an emerging technology based on fundamental principles of quantum mechanics \cite{Nielson2010IntroductionInformation}. Although in its nascent stages, QC research over the last two decades has shown potential for several practical applications \cite{khang2024applications}. This includes a wide range of problems such as factoring, search, material \& chemical simulations, optimization, etc. \cite{Bova2021CommercialComputing}. A quantum algorithm can be modeled and simulated into a physical quantum system either through continuous-analog evolution or, using a sequence of discrete-digital operations in a quantum circuit \cite{Nielson2010IntroductionInformation}. Figure. \ref{fig:program} shows a quantum circuit for GHZ state preparation algorithm having CNOT and Hadamard gate operations in a specific sequence \cite{Nielson2010IntroductionInformation}. Where, qubits are the smallest units of information in quantum computing, powered by properties such as superposition and entanglement. 

It is known that the resources available on earth is limited \cite{Schmidt-Bleek2011TheIntervention}, and due to its unregulated exploitation we'll soon be left with nothing! This is also applicable to the unprecedented use of natural resources for computing applications \cite{Dayarathna2016DataSurvey}. So, it is only reasonable to develop strategies for efficient and sustainable use of available resources. Motivated by this we surveyed and identified that the resource consumption for quantum computers are also very high \cite{Jaschke2023IsAdvantage}. This work doesn't propose a direct solution to optimize for their natural resource consumptions, rather an indirect approach is taken. 

With current hardware being noisy and prone to error \cite{Preskill2018QuantumBeyond}, it becomes difficult to execute large scale problems on a standalone device. So, to effectively utilize the current state of these systems, it becomes important to form a distributed workflow for such large scale problems, and thereby solve them on a distributed environment. Hence, distributed quantum computing has got some traction over the past few years \cite{Caleffi2024DistributedSurvey}. Different approaches have been proposed to formalize such distributed systems\ \cite{Barral2025ReviewComputing} and program execution \cite{Buhrman2003DistributedComputing, Childs2025QuantumConquer, Bisicchia2023DistributingByShots}. Some of these uses classical communication between devices \cite{Bisicchia2023DistributingByShots, Piveteau2024CircuitCommunication}, while other assume quantum communication \cite{Loke2022FromOverview, Qiao2022QuantumChallenges}. 

\begin{table*}[h]
    \centering
\caption{Related works and their comparison with our proposed solution}
\label{tab:related work}
    \begin{tabular}{|>{\centering\arraybackslash}p{0.08\textwidth}|>{\centering\arraybackslash}p{0.08\textwidth}|>{\centering\arraybackslash}p{0.09\textwidth}|>{\centering\arraybackslash}p{0.06\textwidth}|>{\centering\arraybackslash}p{0.08\textwidth}|>{\centering\arraybackslash}p{0.09\textwidth}|>{\centering\arraybackslash}p{0.25\textwidth}|}     
    \hline
 Related work &  Hybrid system &  Distributed workflow &  Error &  Completion time & Network cost & Method used\\\hline
         \cite{Ravi2021AdaptiveCloud}& \ding{51}  & \ding{55} & \ding{51} &  \ding{51} & \ding{55} & Binary linear programming \\\hline
         \cite{Nguyen2024DRLQ:Computing}&  \ding{51}  & \ding{55} & \ding{55} &  \ding{51} & \ding{55} & Deep reinforcement learning \\\hline
         \cite{GiortamisOrchestratingQonductor}&  \ding{51}  & \ding{55} & \ding{51} &  \ding{51} & \ding{55} & Genetic algorithm \\\hline
        \cite{Luo2025AdaptiveLearning} &  \ding{51}  & \ding{55} & \ding{51} &  \ding{51} & \ding{55} & Reinforcement learning \\\hline
         \cite{Zhou2025CloudQC:Computing}&   \ding{55}  & \ding{55} & \ding{55} &  \ding{51} & \ding{51} & Community detection in graphs \\\hline
        \cite{Sane2025OptimizingApproach}&   \ding{55}  & \ding{55} & \ding{55} &  \ding{55} & \ding{51} &  Game theory\\\hline
        \cite{Bahrani2024ResourceNetworks}&   \ding{51}  & \ding{55} & \ding{55} &  \ding{55} & \ding{51} &  Mixed integer linear programming\\\hline
         Ours& \ding{51} & \ding{51} & \ding{51} & \ding{51} & \ding{51} & Graph isomorphism based matching\\ \hline 
\end{tabular}
\end{table*}

As mentioned in \cite{Boschero2025DistributedAndChallenges, Cacciapuoti2020QuantumComputing}, there are several challenge and open questions involving such systems, including but not limited to, networking, optimal program distribution and qubit mapping,  efficient resource allocation and scheduling, etc. This work looks into the problem of efficient resource allocation on a distributed quantum system. We assume  a hybrid quantum-classical network \cite{Phillipson2023ClassificationComputing} and abstract out the the need to distribute a single program, which is different from most of the work in \cite{Luo2025AdaptiveLearning, Zhou2025CloudQC:Computing, Sane2025OptimizingApproach}. Implementation source code can be found at https://github.com/z-ax-qsc/RADIQS. Major contributions of this work include the following:

\begin{enumerate}
    \item We propose the concept of a distributed quantum workflow having an interlinked process of individual quantum programs.
    \item We use all the relevant cost metrics that defines current systems, such as execution time, errors, and communication cost, together for the resource allocation problem.
    \item We propose heuristic-based and randomized algorithms to solve the problem, and perform comprehensive experiments to evaluate them against baselines and state-of-the-art algorithms.
\end{enumerate}

The rest of the paper is organized as follows: Section \ref{sec: related} presents an overview of the existing work related to the similar problem domain as ours. Section \ref{sec: system model} presents a detailed model of the designed system, optimization problem and the proposed algorithm. Section \ref{sec: performance} reports about the experimental setup and the analysis of results. Finally, some concluding remarks and future opportunities are provided in Section \ref{sec: con}.








\section{Related Work}
\label{sec: related}

Resource allocation has been a well known problem for operating systems research \cite{Peterson1985OperatingConcepts, Singh2017ASystems}. It is also applicable to distributed architectures \cite{Goscinski1990ResourceSystems,Krauter2002AComputing, Endo2011ResourceChallenges}, cloud based systems \cite{Jennings2014ResourceChallenges, Professor2012AComputing} and HPC environments \cite{Hussain2013ASystems, Qureshi2020ASystems}. It is  crucial for large-scale computing pipelines to efficiently allocate resources over the available devices.

Current developments towards the new quantum computing paradigm has challenges that need to be addressed specifically for such environments \cite{Corcoles2020ChallengesSystems}. Being a combinatorial optimization problem \cite{papadimitriou1998combinatorial} which can be modeled using some of the common mathematical formulations, see section \ref{sec: problem_def}, it is observed that most of the existing works have also used some of the well known algorithms to solve them, see Table \ref{tab:related work}, with certain modifications specific to quantum systems.

Although, most of the recent work \cite{Ravi2021AdaptiveCloud, Nguyen2024DRLQ:Computing, GiortamisOrchestratingQonductor, Luo2025AdaptiveLearning, Zhou2025CloudQC:Computing} addresses "completion time" in their formulation of resource allocation problem, to the best of our knowledge and as presented in Table \ref{tab:related work}, none of them have tackled for all the three cost parameters, including runtime, error (fidelity) and network costs together. For a robust allocation in the current NISQ era it is important to handle all of these metrics together, and our work strategically combines them in the cost function, see Equation \ref{eq: costfunc}.

With monolithic quantum devices in their early stages of development, in terms of quality and quantity of qubits and gate operations, it is reasonable to study these systems under a distributed setup for practical execution of large-scale problems \cite{Caleffi2024DistributedSurvey}. Where some of the recent work in \cite{Zhou2025CloudQC:Computing, Sane2025OptimizingApproach, Bahrani2024ResourceNetworks} do address this problem within a networked environment, most of them in \cite{Ravi2021AdaptiveCloud, Nguyen2024DRLQ:Computing, GiortamisOrchestratingQonductor, Luo2025AdaptiveLearning} considers a group of standalone quantum devices working collaboratively for the assigned tasks. While \cite{Luo2025AdaptiveLearning} assumes a classical network, \cite{Zhou2025CloudQC:Computing, Sane2025OptimizingApproach, Bahrani2024ResourceNetworks} assumes a quantum network between distributed devices. All of these works assume a single quantum program (circuit) and divide it into multiple chunks for distributed execution. This is different from our use case, where we consider a distributed workflow of standalone quantum programs working in synergy for a larger computational task, see section \ref{sec: distributed system}.

Since the new quantum systems will only supplement existing computing infrastructure and solve large-scale complex problems that are currently intractable, it should be noted that most of the computing tasks will be hybrid in nature \cite{Phillipson2023ClassificationComputing}, where some part is executed on quantum devices and others on classical machines. The authors in \cite{Mantha2024Pilot-Quantum:Management} provide a middleware to manage such problems in a quantum-classical hybrid environment.

\section{System Model and Problem Formulation}
\label{sec: system model}
This section provide details about the formulations of the designed system, optimization problem, constraints, decision metrics, and underlying assumptions for our use case.

\begin{figure*}[h]
    \centering
    \includegraphics[width=0.9\linewidth, height=235pt]{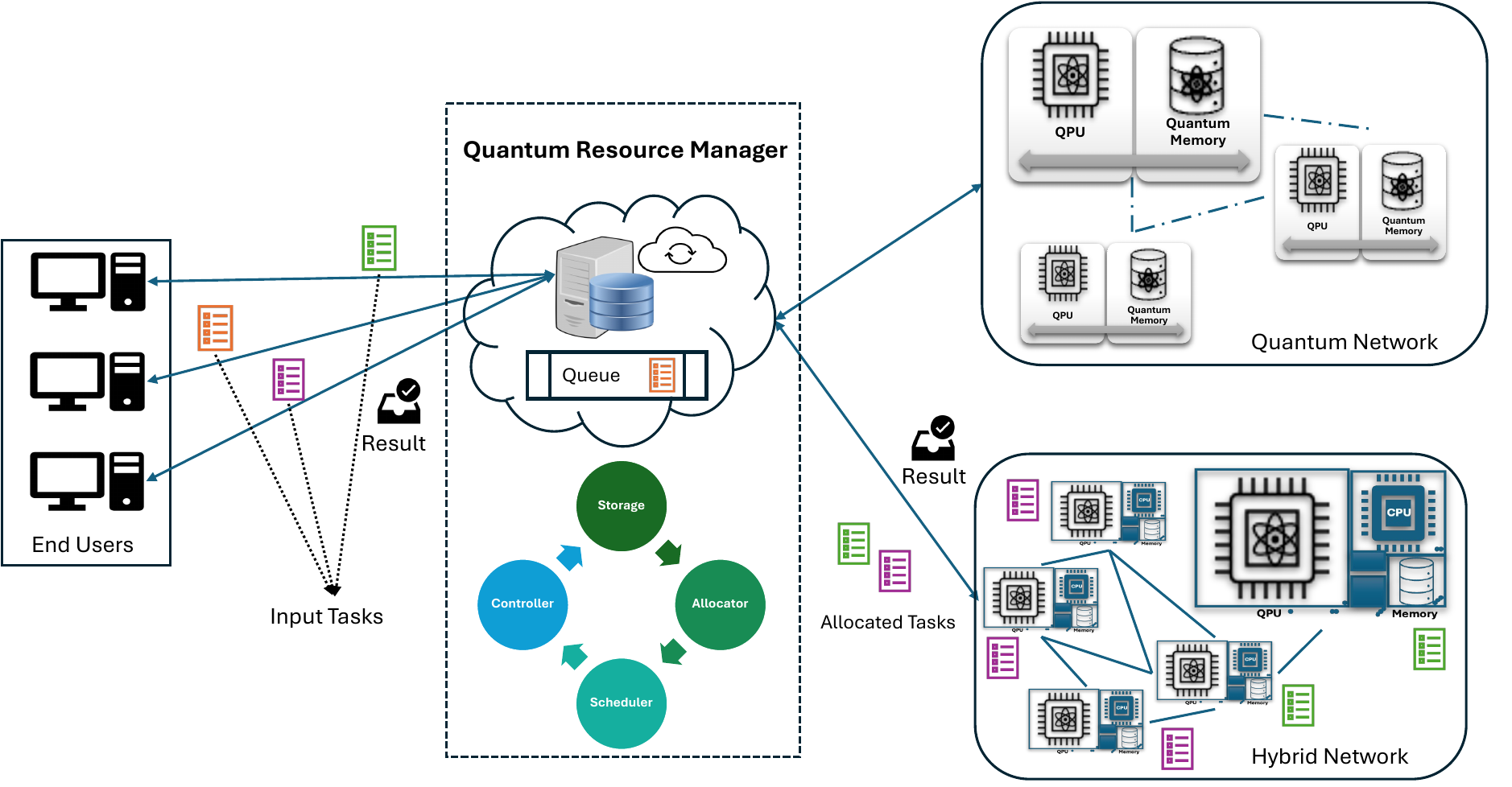}
    \caption{System  Architecture}
    \label{fig:dit_net}
\end{figure*}

\subsection{System Model}
\label{sec: sys model}
We will first introduce the individual components of our system, followed by more details on problem model in the following subsection.
\\


\subsubsection{Distributed quantum system}
\label{sec: distributed system}


  An architecture of a cloud based quantum service manager with distributed quantum resources and multiple clients is shown in Figure \ref{fig:dit_net}. For completeness it also demonstrates a purely quantum interconnect model, but since these are more complex and has not been practically realized, we use the hybrid system for our formulations. Each input task is represented by a quantum workflow as described in sub section \ref{sec: workflow}. The main roles of the service provider (resource manager) include:

\begin{itemize}
    \item A storage to store user requests and computational results.
    \item An allocator module to manage the allocation  of user requests to appropriate devices.
    \item A scheduler to send tasks to quantum devices based on a decided schedule and availability.
    \item A controller to manage the network and other requirements of distributed quantum systems.
\end{itemize}

This work propose new strategies for efficient allocation and assume standard setup for other modules. i.e. as the user requests reach the resource manager, it assigns them optimally to the required set of quantum machines and then schedules them based on their availability. Once a machine processes a user task, the resource manager temporarily stores this result and sends them back to the user once all the tasks in the requested workflow are processed.\\

\subsubsection{Quantum workflow}
\label{sec: workflow}
A quantum workflow $T_i$ is defined by $(G_i,t_i)$, 

Where, \( G_i = ( QC_{ij}, \: E_i ) \) is the graph with nodes representing individual tasks $QC_{ij}$ (quantum circuit) and $E_i$ representing dependency between these tasks. $t_i$ is the arrival time of the workflow request $T_i$, $\forall i \in [1,M] \;\& \; \exists j \in \{2, N\}$. 

Where, M is the total number of workflows and N indicates the maximum allowable task per request $T_i$.

Properties relevant to a task represented by $QC_{ij}$:

\begin{itemize}
    \item $qb_{ij}$ - Number of qubits
    \item $cd_{ij}$ - Circuit depth
    \item $g^{2}_{ij}$ - Number of 2-qubit gates
    \item $mq_{ij}$ - Number of measured qubits
    \item $sht_{ij}$ - Number of shots
\end{itemize}

Examples of sample task workflows are shown in Figure \ref{fig:task graph}. Each example represents a user request with multiple tasks. A simplest use case for a size two workflow (N=2) represents a graph state initialization followed by a QNN task .
\\

\subsubsection{Quantum nodes}

A quantum node $Q_k$ is defined by $(g, e, m, qb, rt)$, where:
\begin{itemize}
    \item $g_k$ - Set of local gate operations
    \item $e_k$ - Error model including; readout error ($e^r_k$), average 2-qubit error ($e^2_k$), average 1-qubit error ($e^1_k$).
    \item $m_k$ - Qubit connectivity at hardware level
    \item $qb_k$ - Number of qubits
    \item $rt_k$ - Runtimes including; $T1_k$ coherence time, $T2_k$ coherence time, average single qubit gate runtime ($rt^1_k$), average two qubit gate runtime ($rt^2_k$) and readout time ($rt^r_k$)
    \item $d1cps_k$ - Depth-1 circuit layer operations per second.
    \item $nat_k$ - Next available time.
\end{itemize}

The set of available QPUs (Quantum Processing Units) is defined by $\mathcal{Q} = \{Q_k\}, \; \forall k \in [1,K]$. The network of available quantum nodes $\mathcal{Q}$ is defined using an undirected graph $G^\mathcal{Q} = (\mathcal{Q}, E^\mathcal{Q})$, where links are predefined using a hybrid quantum-classical network.

\subsection{Problem definition}
\label{sec: problem_def}

Having described the low level components of the designed system, this subsection formally lays out the optimization formulation for the resource allocation problem. 

Let, at a given simulation time (t) the task set chosen for allocation is defined by $\mathcal{T}^t = \{T_i \,| \, t_i <= t \}$ and the available QPUs by the set $\mathcal{Q}$, then our objective is to find an injective mapping $\mathcal{F}: \mathcal{T}^t \to \mathcal{Q}$, where
\begin{equation}
     \mathcal{F}(T_i) = \mathcal{Q}_i=\{(QC_{i1}, Q_x), \, (QC_{i2}, Q_y), ... \}_j
\end{equation}
such that, $\forall \, T_i \in \mathcal{T}^t  \to \mathcal{Q}_i \subseteq \mathcal{Q}$, and
\begin{equation} \label{eq: cc}
 |\mathcal{Q}_i|=|T_i| \; \& \; G_i \subseteq  G^{Q_i} \;  \text{(Connectivity constraint)}
\end{equation}
\begin{equation} \label{eq: qc}
   ( \forall Q_k \in \mathcal{Q}_i \; \& \; QC_{ij} \in T_i) \to qb_{ij} \leq qb_k  \;  \; \text{(Qubit constraint)}
\end{equation}
and, minimize the weighted cost function
\begin{equation}
 min \bigg(\zeta A_{i} + (1-\zeta)(\alpha E_{i} + \beta R_{i} + \gamma N_{i}) \bigg), \; \forall T_i \in \mathcal{T}^t
 \label{eq: costfunc}
\end{equation}
such that, $\forall Q_k \in \mathcal{Q}_i \; \& \; QC_{ij} \in T_i$,\\
\begin{itemize}
    \item Next available time \((A_{i}) = \text{max}(\{nat_k\})\)\\
    \item Total error \( (E_{i}) = \sum_{j,k}\mathcal{E}(QC_{ij} \, ,\, Q_k )\)\\
    \item  Total runtime \((R_{i}) = \sum_{j,k}\mathcal{R}(QC_{ij} \, ,\, Q_k )\)\\
    \item Total network cost \((N_{i}) = \mathcal{N}(T_i \, ,\, \mathcal{Q}_i )\)
\end{itemize}

Where, weights $\zeta \leq 1$, $\alpha + \beta + \gamma = 1$ and $\mathcal{E}, \mathcal{R}, \mathcal{N}$ measures the cost associated to the execution of assigned task on the chosen machine. It is to be noted that the constraints in equations \ref{eq: cc}, \ref{eq: qc} indicate that the assigned quantum machines should have enough qubits to execute the program and the connectivity between these QPUs should match that of requested workflow.

\begin{figure}[h]
    \centering
    \includegraphics[width=0.9\linewidth, height=200pt]{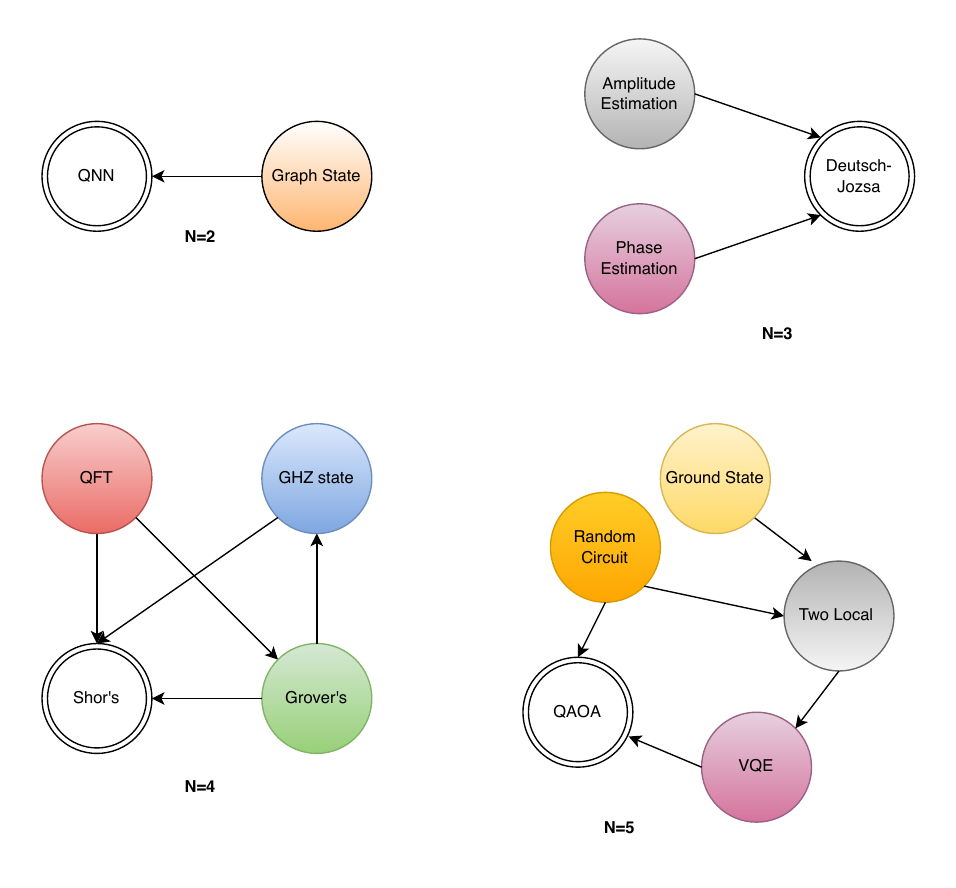}
    \caption{Sample quantum workflows of different sizes, where 'N' denote the number to tasks in each workflow.}
    \label{fig:task graph}
\end{figure}

\subsection{System Cost Metrics}
Sub section \ref{sec: problem_def} have already hinted about the three main cost functions, namely $\mathcal{E}, \mathcal{R}, \mathcal{N}$. Here, we provide more details on their definition and mathematical formulations.

\subsubsection{\textbf{Error}}
The error of executing a task on a quantum computer is defined by properties of both the participating entities. The following formulation is constructed, derived and taken from the estimates and work presented in \cite{McKay2023BenchmarkingScale}, \cite{Magesan2012CharacterizingBenchmarking}. It is essentially a combination of errors associated with 1,2-qubit gate operations and the readout error for executing the quantum circuit $QC_{ij}$ onto the quantum machine $Q_{k}$.

We define, 
\begin{multline}
\label{eq: error}
 \mathcal{E}(QC_{ij} \, ,\, Q_k )  =  \\
 1 - (1 - e^1_k)^{cd_{ij}} * (1 - e^2_k)^{\sqrt{g^2_{ij}}} * (1 - e^r_k)^{qb_{ij}}
\end{multline}

\subsubsection{\textbf{Runtime}}
As mentioned in \cite{WackQualityComputers}, the runtime of a quantum program can be characterized by its quantum volume and the speed of the quantum computer usually defined by CLOPS (Circuit Layer Operations Per Seconds). Since the current NISQ devices are error prone, it is also important to consider the number of shots for statistical significance.

We use a simplified version of the formulation described in \cite{WackQualityComputers}, using the circuit depth and D1CPS(Depth-1 Circuit Layer Operations Per Second) as a proxy of quantum volume and CLOPS, respectively. s.t.
\begin{equation}
 \mathcal{R}(QC_{ij} \, ,\, Q_k )  =  \frac{cd_{ij} * sht_{ij}}{d1cps_k} 
\end{equation}

\subsubsection{\textbf{Communication latency}}
As mentioned earlier, we assume a hybrid quantum-classical network between quantum devices. Such that, 
\begin{multline}
\label{eq: comm}
\mathcal{N}(T_i \, ,\, \mathcal{Q}_i ) =  \\
    \sum_{(j,j') \in E_i \: \& \: (k,k') \in E^{\mathcal{Q}_i} } 
  N^q(j,j',k,k') + N^c(j,j',k,k') \\
  s.t. \\
  N^q(j,j',k,k')  =  \frac{\mathcal{N}^q(QC_{ij} \, ,\, Q_k) + \mathcal{N}^q(QC_{ij'} \, ,\, Q_{k'})}{2} \\
  \& \\
   N^c(j,j',k,k')  = \frac{\mathcal{N}^c(QC_{ij} \, ,\, Q_k) + \mathcal{N}^c(QC_{ij'} \, ,\, Q_{k'})}{2}
\end{multline}

Where, $N^q(j,j',k,k')$ is the cost of quantum link between nodes $Q_k$ and $Q_{k'}$ executing tasks $QC_{ij}$, $QC_{ij'}$, respectively. Similarly,  $N^c(j,j',k,k')$ is the cost of classical link.

As proposed in some of the recent works, such as in \cite{Pompili2021RealizationQubits} , \cite{Shi2020ConcurrentDesigns}, \cite{Beukers2024Remote-EntanglementInterfaces}, quantum networks are mainly realized using EPR (Einstein-Podolsky-Rosen) pair generation and BSM (Bell state measurements) over interconnected nodes with single or multi-hop transfer links (switches). Based on this,

\begin{equation}
    \mathcal{N}^q(QC_{ij} \, ,\, Q_k) = \frac{\varrho * 10 * qb_{ij} * rt^2_k }{HM(T1_k, T2_k) * \eta^{n_s}}
\end{equation}
where, $\varrho$ is connection success probability, $\eta$ is transmission efficiency, $n_s$ number of switch between links, $HM(T1_k, T2_k)$ is harmonic mean of the two coherence times.

Whereas, the characteristics of the classical information transfer are derived from the work done in \cite{CarreraVazquez2024CombiningCommunication} and \cite{ac1979some} for an interconnected  system of distributed quantum machines. s.t.,

\begin{equation}
    \mathcal{N}^c(QC_{ij} \, ,\, Q_k) = \rho * mq_{ij}
\end{equation}
Where, $\rho$ is the classical communication latency. 

\subsection{Key Assumptions}
This work assumes the following properties about the designed system:
\begin{enumerate}
    \item Quantum machines of the same qubit modality (superconducting) but with different system characteristics is used.
    \item All circuits are mappable to any given hardware topology and their transpiled versions are used for allocation problem.
    \item Cost for retention of information in quantum/classical memory is ignored.
    \item Task graphs are represented by directed acyclic graphs for ordered execution. See Figure. \ref{fig:task graph} for sample 
    \item Sub tasks execute asynchronously under hybrid quantum -classical communication.
    \item All links between any pair of quantum devices are assumed to be similar in terms of underlying networking modules used.
    \item There are enough machines to cater any individual task requirements. But, it should be noted that joint requirements of a workflow is dependent on connectivity of the distributed system.
    \item Each QPU has their own queue and tasks allocated to them are executed in FCFS manner.
\end{enumerate}

\begin{figure}[h]
    \centering
    \includegraphics[width=.7\linewidth, height = 350pt]{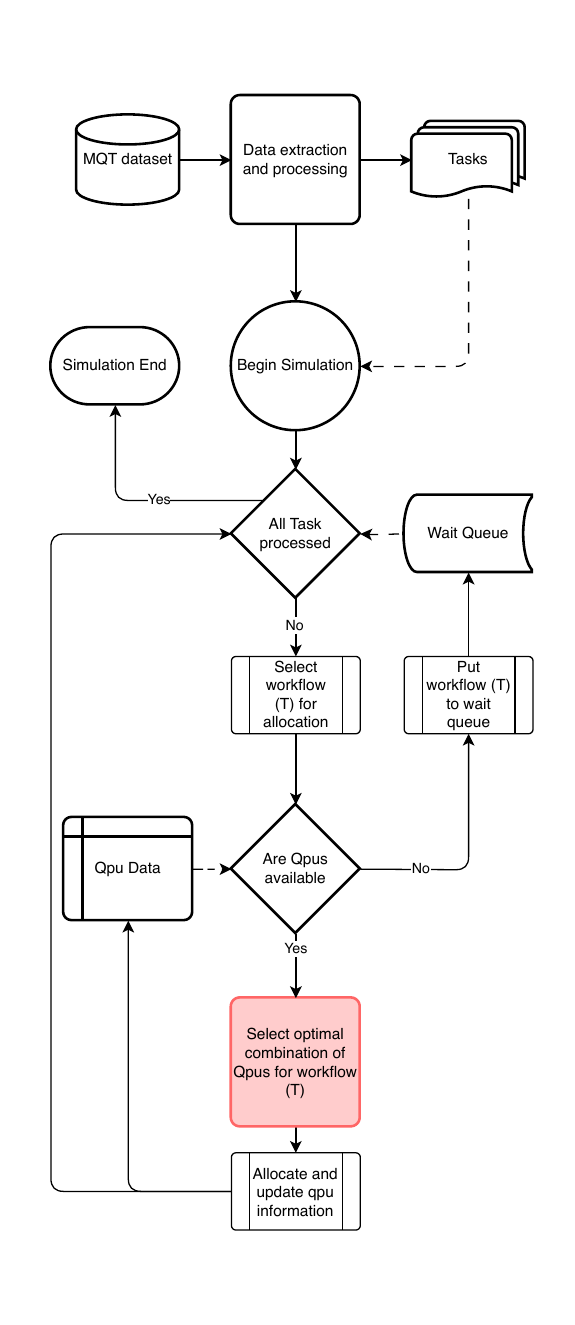}
    \caption{Process pipeline for experimental simulation}
    \label{fig:pipeline}
\end{figure}

\subsection{Proposed Algorithm}

\begin{algorithm}[h]
\caption{SoftIso}\label{alg:softiso}

\KwIn{$T_i \text{(chosen workflow)}, G^\mathcal{Q} \text{(resource network)}$}
\KwOut{$\mathcal{Q}_i \text{(assigned QPUs)}$}
\vspace{5pt}
$ iter \gets GraphMatcher(G^\mathcal{Q}, G_i) $ \tcp{\cite{Cordella2004AGraphs, Houbraken2014TheEnumeration}}
$mincost \gets \infty$\;
$maxcost \gets -\infty$\;
$prevcost \gets 0$\;
$counter \gets 0$\;
\vspace{5pt}
\tcp{$G^\mathcal{Q_s} \subset G^\mathcal{Q}$}
\For{$G^\mathcal{Q_s} \in iter$}{
    \tcp{with workflow i and QPU set $\mathcal{s}$}
    $cost \gets \zeta A_{\mathcal{s}} + (1-\zeta)(\alpha E_{\mathcal{s}} + \beta R_{\mathcal{s}} + \gamma N_{\mathcal{s}})$\;
    $maxcost \gets \text{max}(cost, maxcost)$\;
  \If{$cost < mincost$}{
    \tcp{Conditions as in Eq. \ref{eq: cc}, \ref{eq: qc}}
    \If{$\text{mapping}(G^\mathcal{Q_s}, G_i) \; \& \; \text{qubits}(\mathcal{Q_s}, T_i)$}{
        $mincost \gets \text{min}(cost, mincost)$\;
        $\mathcal{Q}_i \gets \mathcal{Q_s}$\;
    }

    \tcp{Early stopping}
    \If{($|\Delta(cost, maxcost)|> THRES\_MAX$ \textbf{AND} $|\Delta(cost, prevcost)|> THRES\_PREV$) \textbf{OR} $counter \geq 10^{|T_i|}$ }{
        break\;
    }
  }
    $counter \gets counter +  1$\;
}
\end{algorithm}

As depicted in Figure \ref{fig:pipeline}, the simulation happens in discrete time intervals, where at a give simulation time (t) it collects all the workflows submitted by various users and pick them one-by-one on FCFS basis for passing it to the assignment algorithm for allocation. In case there are multiple user requests at the same time instance, it choose workflows in ascending order of qubit requirements or break tie based on user priority. This process continues until all tasks are assigned. The algorithms described in \ref{alg:softiso}, \ref{alg:random} is used to assign one workflow at a time having multiple sub tasks.

To meet the objectives defined in sub section \ref{sec: problem_def}, we propose \textbf{SoftIso}, an extended version of graph isomorphism algorithms in \cite{Cordella2004AGraphs, Houbraken2014TheEnumeration} with early stopping criteria for non-exhaustive search process, see Algorithm \ref{alg:softiso}. This involve deviation thresholds from the maximum cost value and the previous cost value as THRES\_MAX and THRES\_PREV, respectively. The "\textit{mapping}" and the "\textit{qubits}" functions evaluates the conditions in Equations. \ref{eq: cc} and \ref{eq: qc}, respectively.  

Since graph isomorphism is a NP hard problem, our solution with early stopping criteria provide reasonable guarantees for a faster solution on average. It is to be noted that it will still have exponential overhead for large and complex problem in worst case. Results in Table \ref{tab:performance} validates this with acceptable tradeoffs on system cost values. 

\begin{algorithm}[h]
\caption{RandomAware}\label{alg:random}
\KwIn{$T_i \text{(chosen workflow)}, G^\mathcal{Q} \text{(resource network)}$}
\KwOut{$\mathcal{Q}_i \text{(assigned QPUs)}$}
\vspace{5pt}

$mincost \gets \infty$\;
\vspace{5pt}

\While{$trials \leq |T_i|$}{
    \tcp{Initialize with empty set}
    $\mathcal{Q}_s \gets \Phi$\;

    \tcp{sort tasks by qubits}
    $T_i \gets \text{sort($T_i, \; qb_{ij}$)}$\; 
    
    \tcp{Find task to QPU mapping}
    \For{$QC_{ij} \in T_i$}{
        $S_j \gets \{Q_k \, | \, qb_k \geq qb_{ij}\}$\;
        $\mathcal{Q}_s  \gets \mathcal{Q}_s  \cup \text{RandomSelect}(S_j)$\;
    }

    \tcp{with workflow i and QPU set s}
    $cost \gets \zeta A_{s} + (1-\zeta)(\alpha E_{s} + \beta R_{s} + \gamma N_{s})$\;
  
  \tcp{Condition check as in Equation. \ref{eq: cc}}
  \If{$cost < mincost \; \textbf{and} \; \text{mapping}(G^{\mathcal{Q}_s} , G_i) $}{
        $mincost \gets \text{min}(cost, mincost)$\;
        $\mathcal{Q}_i \gets \mathcal{Q}_s $\;
  }
    $trials \gets trials +  1$\;
}
\end{algorithm}

In addition to this we have also designed a randomized selection algorithm, \textbf{RandomAware}, as described in Algorithm \ref{alg:random}. This algorithm assign tasks in ascending order of qubits to a random QPU having at most that many qubits. This is done for multiple trials and the final allocation is chosen from the one having lowest cost value among all the feasible trials. The number of random trials is set to the number of tasks in the input workflow. Although this algorithm doesn't perform best in all the scenarios, it can still be beneficial over other algorithms for some use cases, see section \ref{sec: res} for more detail.

\section{Performance Evaluation}
\label{sec: performance}

We conduct multiple experiments to analyze the performance of proposed algorithms for different scenarios and evaluation metrics. Details about the common experimental setup and the parameters used are provided in subsection \ref{sec: setup}.

We report the comparisons with the procedure used in CloudQC \cite{Zhou2025CloudQC:Computing}, and a baseline greedy strategy named GreedyDfs. GreedyDfs assign tasks based on ascending order of qubit requirements to the nodes in ascending order of qubits availability and depth-first-search order of node connectivity. Our results in section \ref{sec: res} indicate promising improvements over both of these techniques.  


\subsection{Experiment Setup}
\label{sec: setup}
An open source discrete-event-based quantum simulator, QSimPy \cite{Nguyen2025QSimPy:Management}, is used for simulation experiments. It is extended as per our use case with elements including virtual representations of quantum workflows, tasks and nodes. We use calibration data from actual hardware and quantum task definition from benchmarking datasets.\\


\subsubsection{Dataset}
There are two main data sources for our simulations, one for the quantum node properties and the other for quantum task definitions.

\paragraph{\textbf{Nodes}}
Calibration data for three IBM machines, namely brisbane, torino and marrakesh, are loaded from \cite{ibm-cal}. Details of these machines can be found in Table \ref{tab:nodes}. Every simulation experiment randomly chooses the required number of nodes, out of these three choices, and decide a random topology based on network success probability $\varrho$. 


\paragraph{\textbf{Tasks}}
Quantum circuit definitions from MQT Bench dataset \cite{quetschlich2023mqtbench} is used, with properties as mentioned in Table \ref{tab:tasks}. Loaded circuits are compiled using qiskit transpiler for the three target machines. Quantum workflows of sizes ranging from 1-5 is created by randomly choosing them from a set of feasible combinations. 
For every simulation experiment an initial collection of user workflows is created, we call them workload, using Poisson's distribution for arrival times and an uniformly sampled collection of program circuits with qubits between 1 and 100.

\begin{table}[h]
    \centering
    \caption{Node properties}
    \label{tab:nodes}
    \begin{tabular}{|>{\centering\arraybackslash}p{0.35\linewidth}|>{\centering\arraybackslash}p{0.15\linewidth}|>{\centering\arraybackslash}p{0.15\linewidth}|>{\centering\arraybackslash}p{0.15\linewidth}|}
    \hline
        \textbf{Properties} & \textbf{IBM-brisbane} & \textbf{IBM-torino} & \textbf{IBM-marrakesh}\\\hline
        Number of qubits & 127& 133& 156\\\hline
        CLOPS & 180000 & 220000 & 200000\\\hline
        One qubit runtime (s) & 60e-9 & 32e-9 & 36e-9\\\hline
        Two qubit runtime (s) & 660e-9 & 68e-9 & 68e-9\\\hline
        Readout runtime (s) & 1600e-9 & 1560e-9 & 2584e-9\\\hline
        T1 coherence time (s) & 220.53e-6 & 181.41e-6 & 188.11e-6\\\hline
        T2 coherence time (s) & 128.92e-6 & 138.75e-6 & 111.97e-6\\\hline
        Median readout error &2.393e-2 &2.991e-2 &1.074e-2 \\\hline
        Median single qubit error &2.517e-4 &3.296e-4 &3.047e-4 \\\hline
        Median two qubit error &7.042e-3 &2.68e-3 &2.451e-3 \\
    \hline
    \end{tabular}
\end{table}

\begin{table*}[h]
    \centering
    \caption{Task properties}
    \label{tab:tasks}
    \begin{tabular}{c|c}\toprule
        \textbf{Task property} & \textbf{Details} \\\midrule
        Source & MQT Bench \cite{quetschlich2023mqtbench}\\
        Number of qubits in tasks & [5-100] \\
        Program types & Random Circuit, Ground State, QFT, GHZ, QPE, AE, Graph State, Grover, DJ, QAOA, QNN, VQE, Shor's\\
        Workflow size & [1-5] \\
        Compiler & Qiskit \\ 
        Arrival distribution & Poissons \\
        Simulated hardwares & ibm-brisbane, ibm-marrakesh, ibm-torino\\
        \bottomrule
    \end{tabular}
\end{table*}

\subsubsection{Experiment Parameters}
\label{sec: exp_params}

We have assumed a single-switch path between interlinked node ($n_s$ = 1) and the transmission efficiency ($\eta$) of 1 dB. To define the connectivity between nodes, the success probability of quantum link ($\varrho$) is set to 0.5, and the classical communication latency ($\rho$) to 0.02. The network of quantum nodes (resources) is created by randomly choosing the required devices from the set of available machines. We use equal weights for the availability and the three system metrics in Equation \ref{eq: costfunc}. i.e. $\zeta = 0.5 \; \text{and} \; \alpha = \beta = \gamma = 1/3$, for a balanced consideration of all the cost functions. THRES\_MAX and THRES\_PREV are set to 0.1 and 0.03, respectively for the SoftIso algorithm.

It is to be noted that all the values of different cost function evaluations are normalized between 0-1 for an equivalent comparison. The parameters mentioned here are common for most of the results, unless stated otherwise.\\

\subsubsection{Evaluation Metrics}
Experiments are evaluated for six major metrics, including two performance metrics such as algorithm run time and percentage workload completion, and four system metrics like execution time, wait time, fidelity and communication cost. These help us in capturing an overall pictures of the designed system and perform a comprehensive comparison of proposed algorithms under different scenarios. All values are calculated as an average of 100 random experiments with same parameters but different workloads. A workload is defined as a collection of user workflows received by the allocator. Brief description of these metrics include:

\paragraph{\textbf{Execution Time}} It is the total time taken by a given workload, with multiple workflows, to run on the assigned QPUs. This is characterized mainly by the gate runtime of the QPU.

\paragraph{\textbf{Wait Time}} It is the sum of waiting time for all the tasks in their respective QPU queue. It is an important metric to evaluate the  efficiently of allocation algorithms.


\paragraph{\textbf{Fidelity}} Fidelity measures the accuracy of the results generated by a noisy QPU for executing the assigned task. There are different ways to calculate the fidelity of program execution \cite{Nielson2010IntroductionInformation}, and in our case we use the simplest formulation as:
\begin{equation}
    Fidelity = 1- Error
\end{equation}
Where, the error is characterized as defined in Equation \ref{eq: error}. Higher the fidelity, better the results. Here, we measure it as the average fidelity of all the tasks submitted for an experiment.

\paragraph{\textbf{Communication overhead}} Along with the fidelity and the execution time, this metric helps in evaluating the overall system performance of the resource allocation strategy. This is calculated as the sum of total communication cost, see Equation \ref{eq: comm}, for running all the workflows to the assigned network of interconnected QPUs.

\paragraph{\textbf{Algorithm runtime}} It is the total cpu time required to simulate/run an experiment by a given allocation algorithm. The results for this metric are reported in Table \ref{tab:performance}.

\paragraph{\textbf{Task completion}} This measures the percentage of the total tasks that are successfully allocated by an algorithm, over all the tasks in the workload.\\

\begin{figure*}[h]
     \centering
 \begin{subfigure}{0.33\linewidth}
         \centering
         \includegraphics[width=\linewidth, height=100pt]{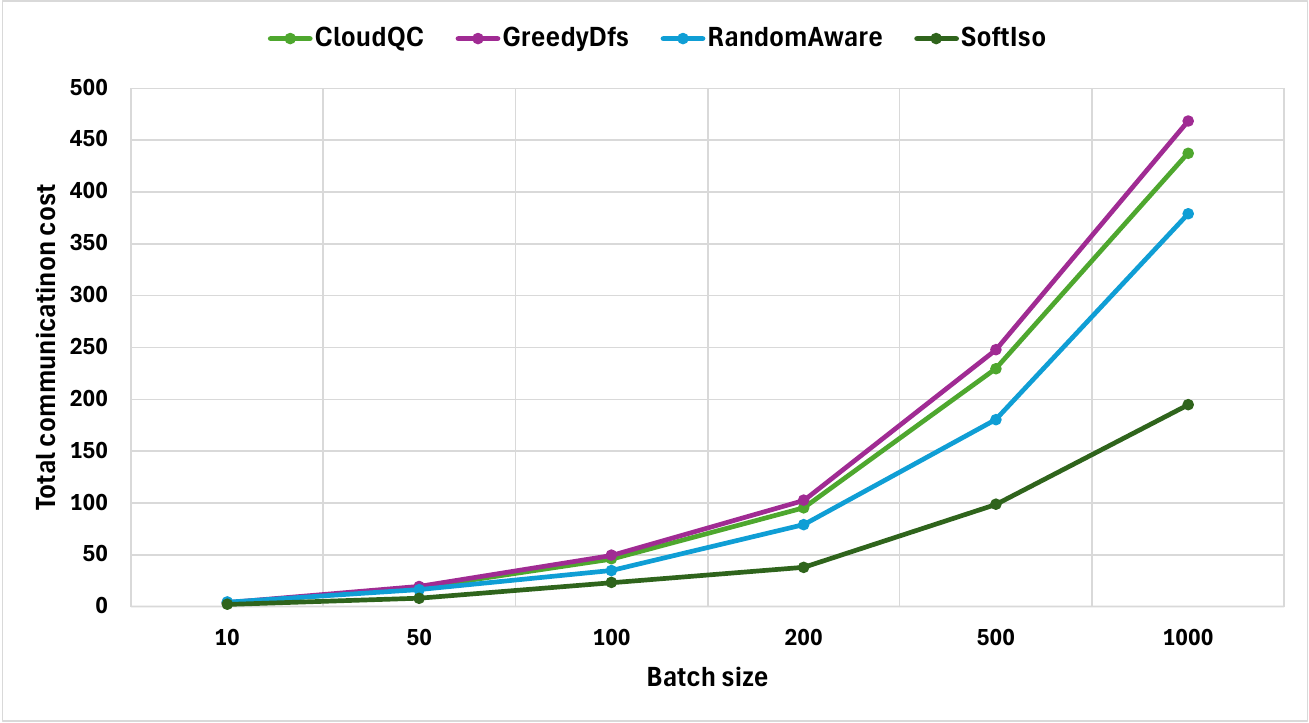}
         \caption{}
         \label{fig:com_batch}
     \end{subfigure}%
     \hfill
     \begin{subfigure}{0.33\linewidth}
         \centering
         \includegraphics[width=\linewidth, height=100pt]{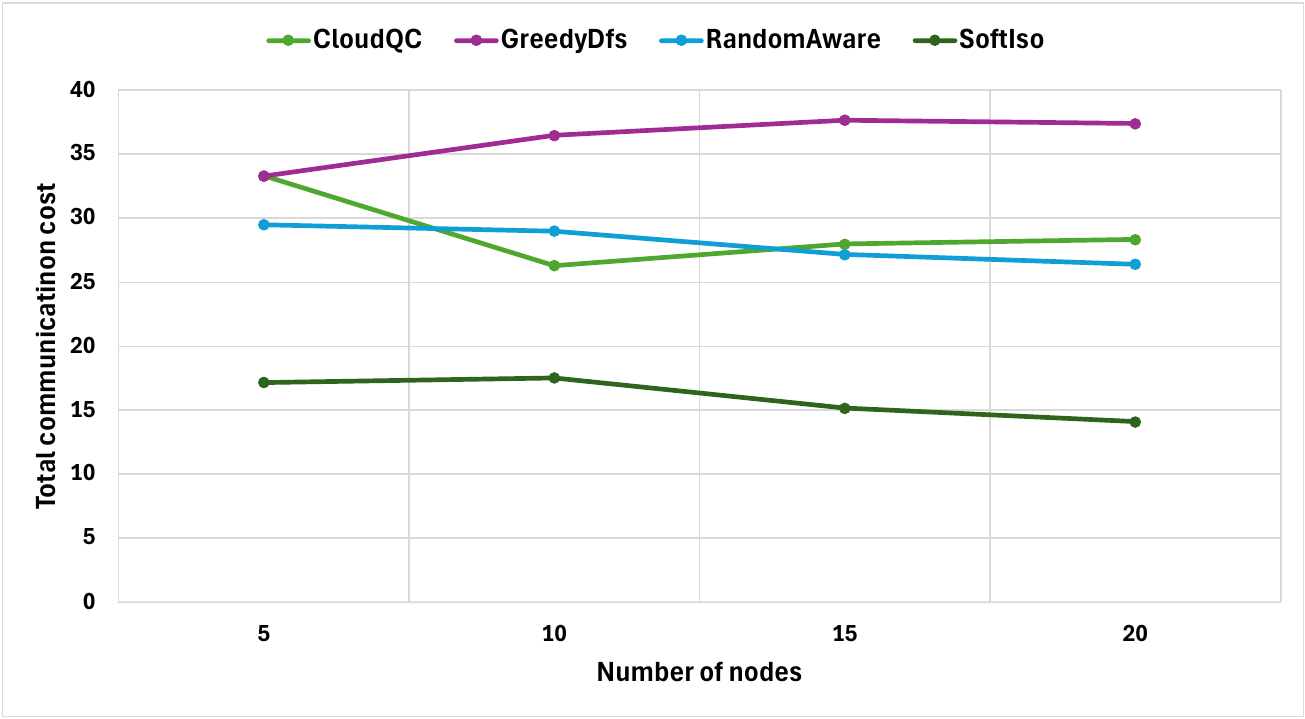}
         \caption{}
         \label{fig:com_node}
     \end{subfigure}%
     \hfill
     \begin{subfigure}{0.33\linewidth}
         \centering
         \includegraphics[width=\linewidth, height=100pt]{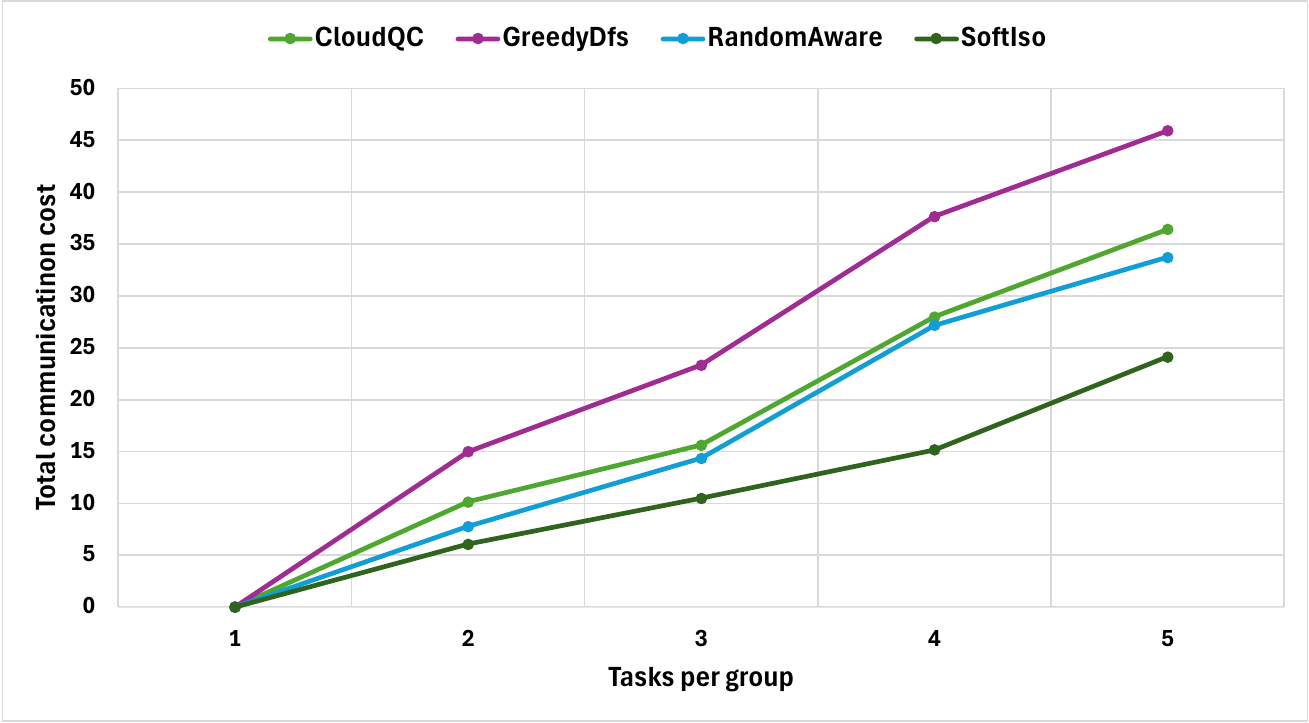}
                  \caption{}
         \label{fig:com_task}
     \end{subfigure}
        \caption{Trends for communication cost by (a) workload size for allocation, (b) number of nodes in the network and (c) distinct tasks per workflow}
        \label{fig:trend_comm}
\end{figure*}

\begin{figure*}[h]
     \centering
 \begin{subfigure}{0.33\linewidth}
         \centering
         \includegraphics[width=\linewidth, height=100pt]{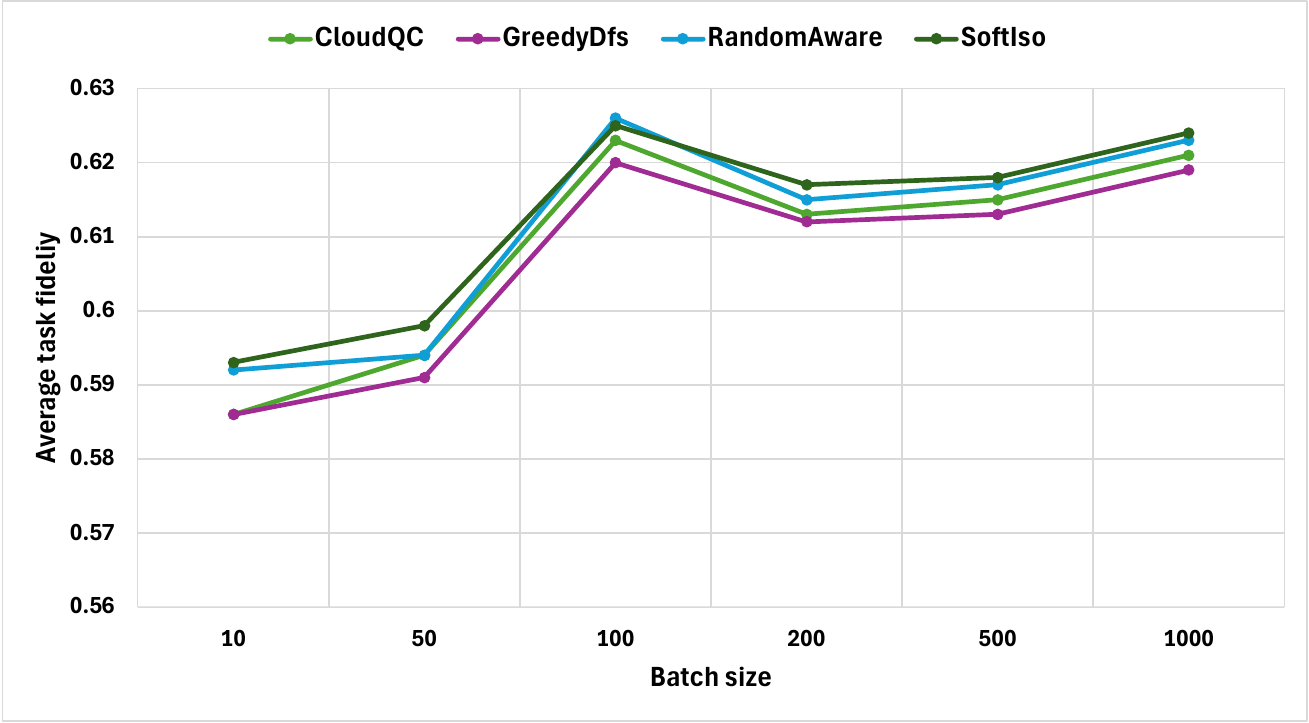}
                  \caption{}
         \label{fig:fid_batch}
     \end{subfigure}%
     \hfill
     \begin{subfigure}{0.33\linewidth}
         \centering
         \includegraphics[width=\linewidth, height=100pt]{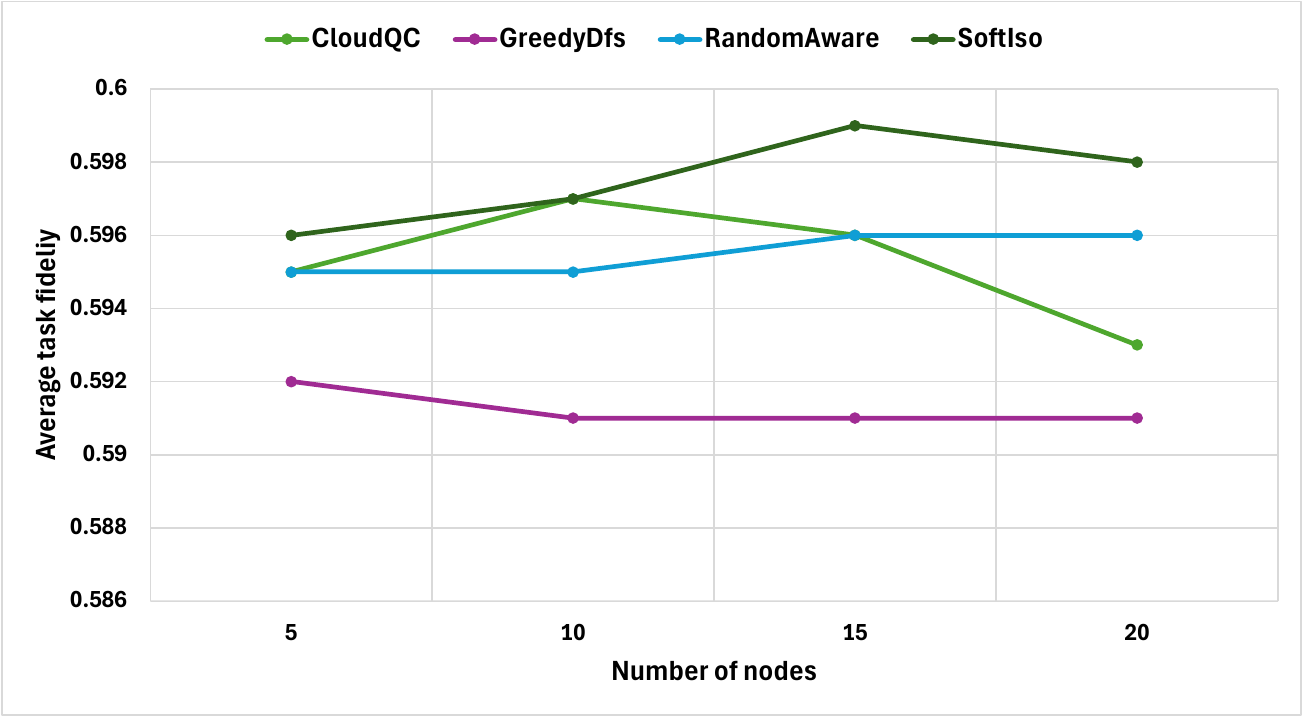}
                  \caption{}
         \label{fig:fid_node}
     \end{subfigure}%
     \hfill
     \begin{subfigure}{0.33\linewidth}
         \centering
         \includegraphics[width=\linewidth, height=100pt]{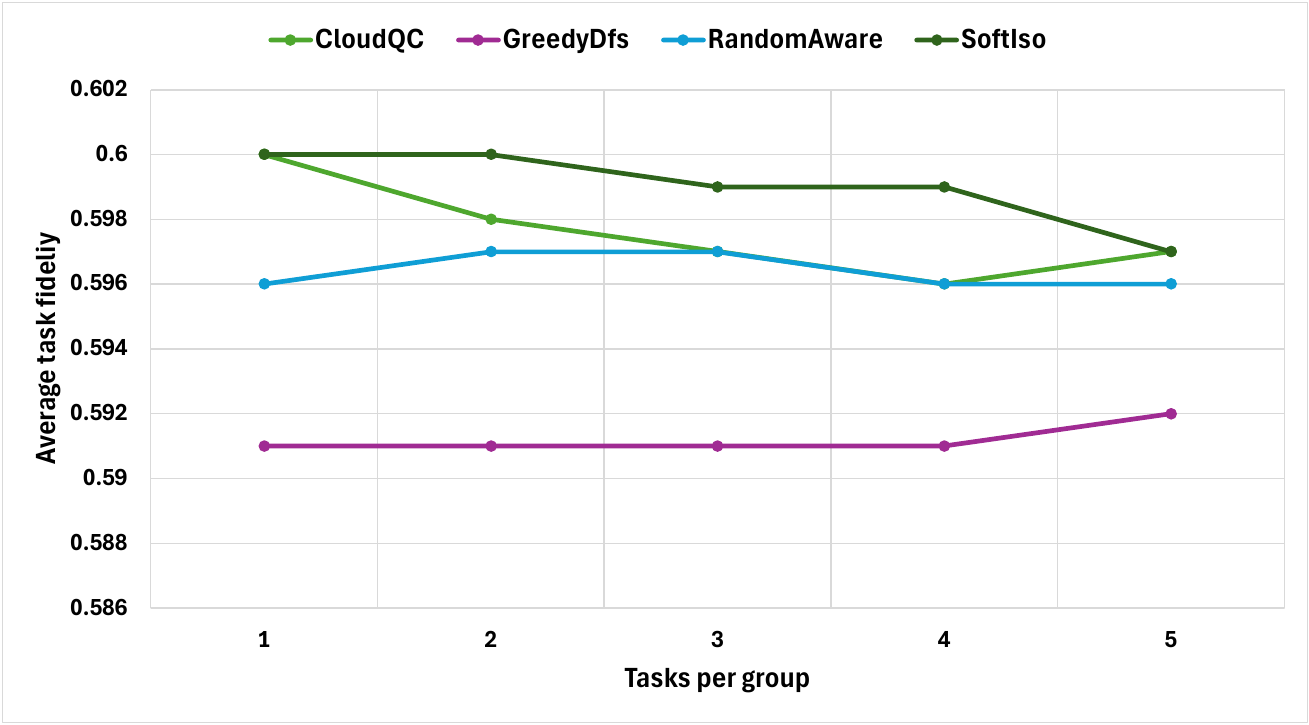}
                 \caption{}
         \label{figfid_task}
     \end{subfigure}
        \caption{Fidelity trends by (a) workload size for allocation, (b) number of nodes in the network and (c) distinct tasks per workflow}
        \label{fig:trend_fid}
\end{figure*}

\begin{figure*}[h]
     \centering
     \begin{subfigure}{0.33\linewidth}
         \centering
         \includegraphics[width=\linewidth]{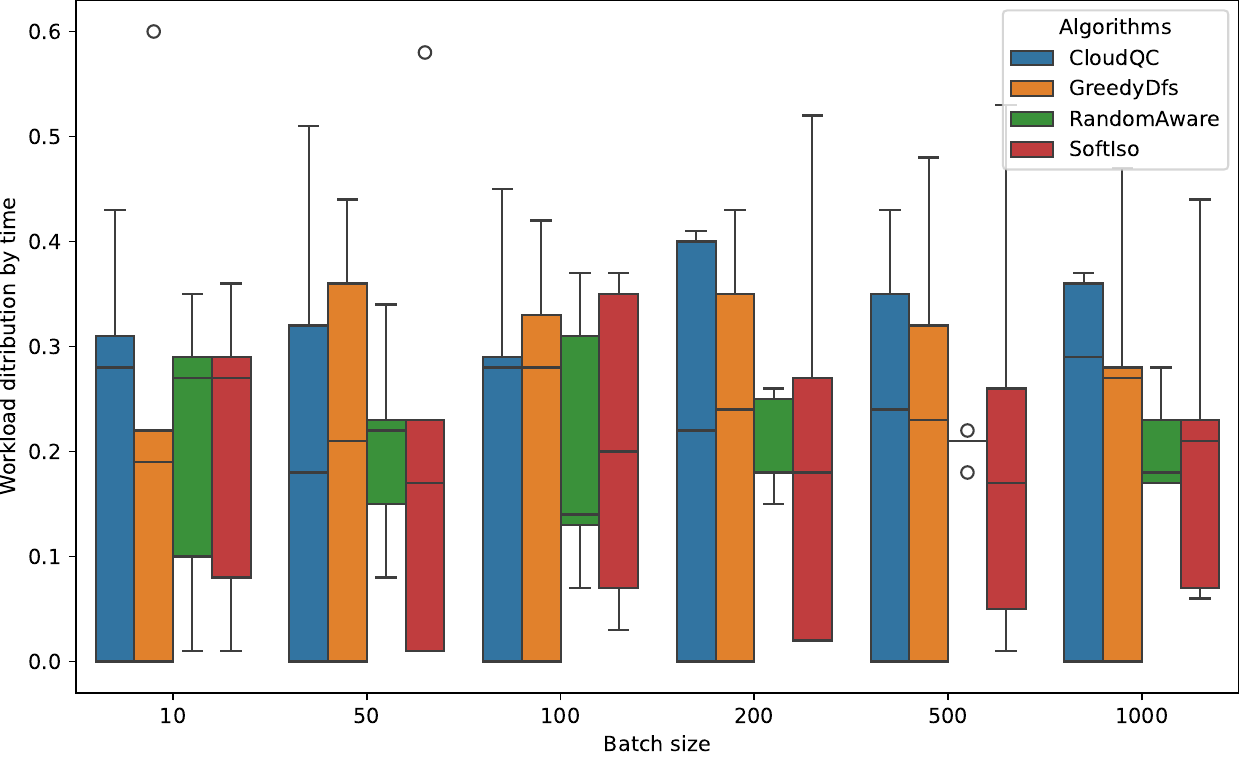}
                  \caption{}
         \label{fig:dt1_batch}
     \end{subfigure}%
     \hfill
     \begin{subfigure}{0.33\linewidth}
         \centering
         \includegraphics[width=\linewidth]{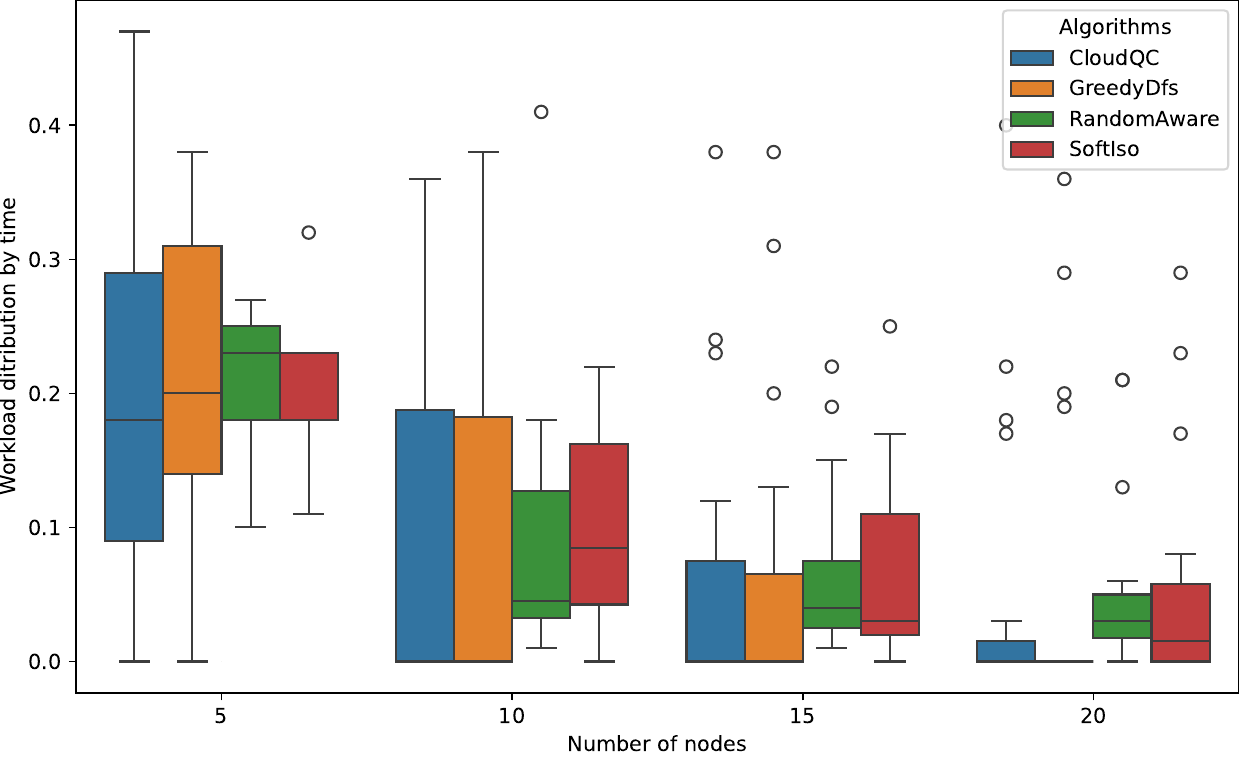}
                  \caption{}
         \label{fig:dt1_node}
     \end{subfigure}%
     \hfill
     \begin{subfigure}{0.33\linewidth}
         \centering
         \includegraphics[width=\linewidth]{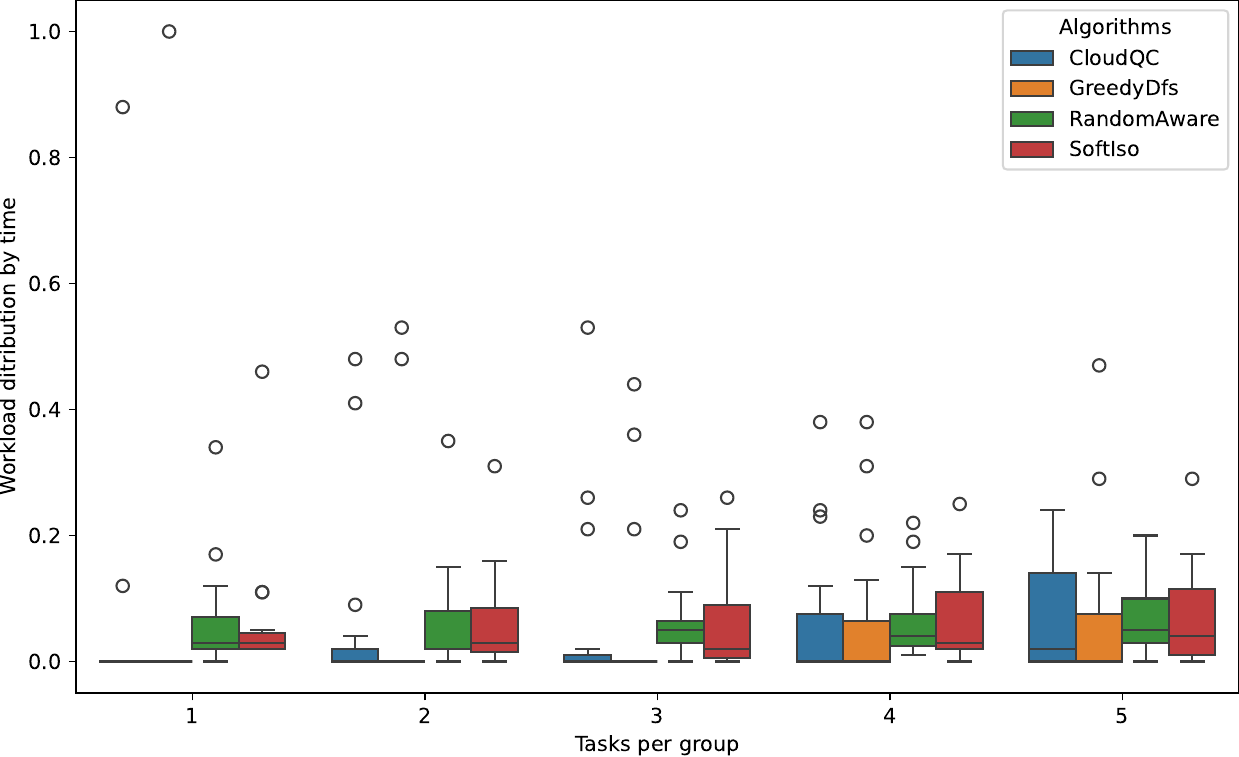}
                \caption{}
         \label{fig:dt1_task}
     \end{subfigure}
        \caption{Time distribution of workload per QPU, for different (a) workload size for allocation, (b) number of nodes in the network and (c) distinct tasks per workflow}
        \label{fig:dist_time}
\end{figure*}

\begin{figure*}[h]
     \centering
     \begin{subfigure}{0.33\linewidth}
         \centering
         \includegraphics[width=\linewidth, height=100pt]{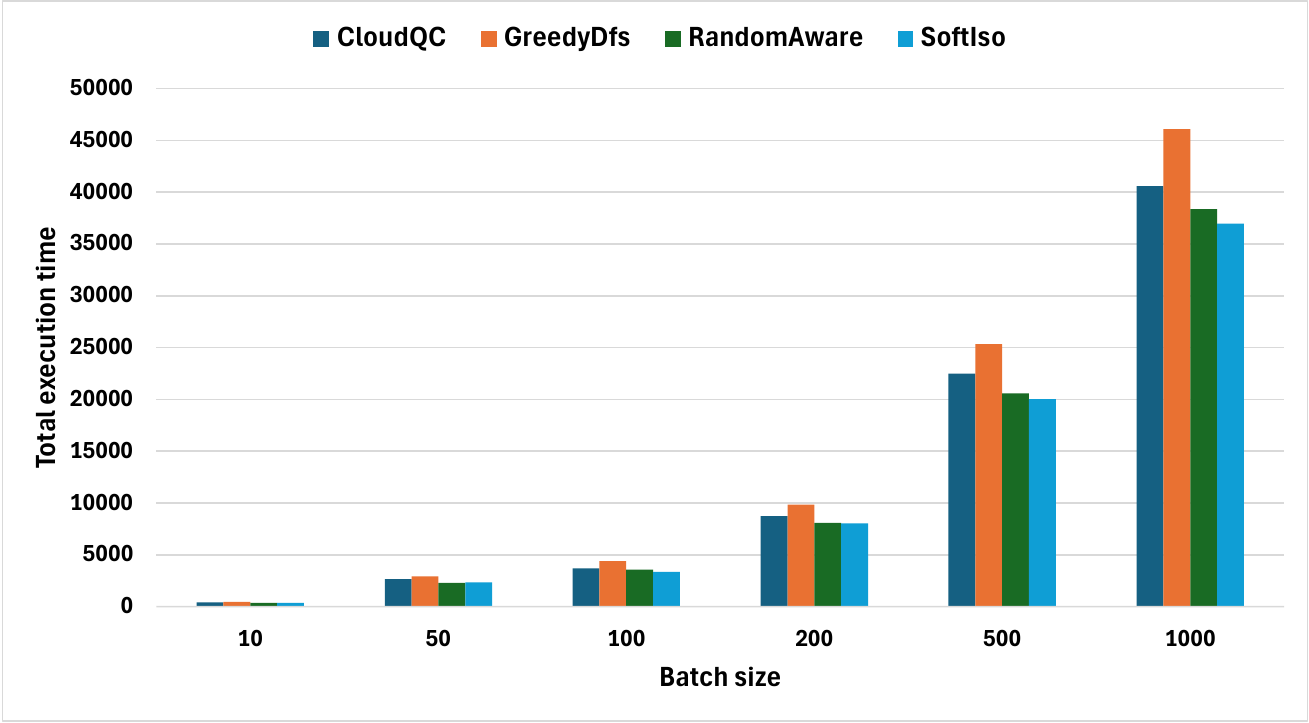}
                  \caption{}
         \label{fig:exec_batch}
     \end{subfigure}%
     \hfill
     \begin{subfigure}{0.33\linewidth}
         \centering
         \includegraphics[width=\linewidth, height=100pt]{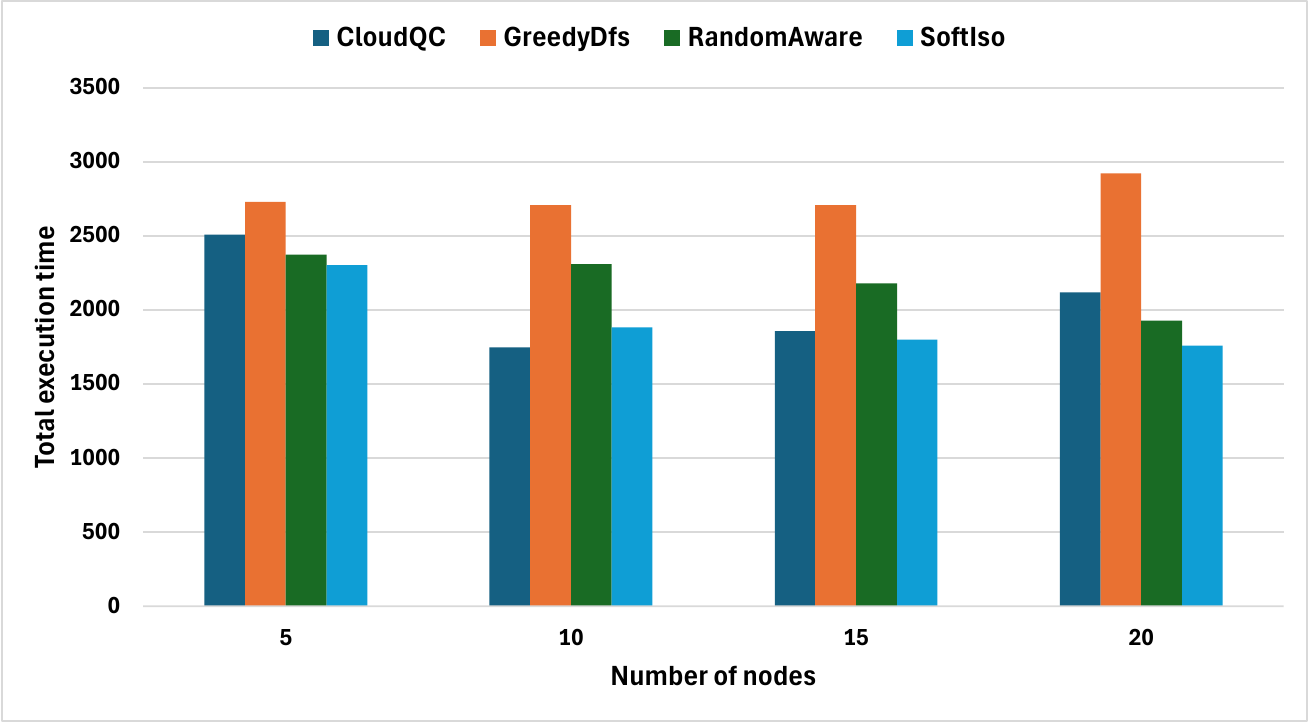}
                 \caption{}
         \label{fig:exec_node}
     \end{subfigure}%
     \hfill
     \begin{subfigure}{0.33\linewidth}
         \centering
         \includegraphics[width=\linewidth, height=100pt]{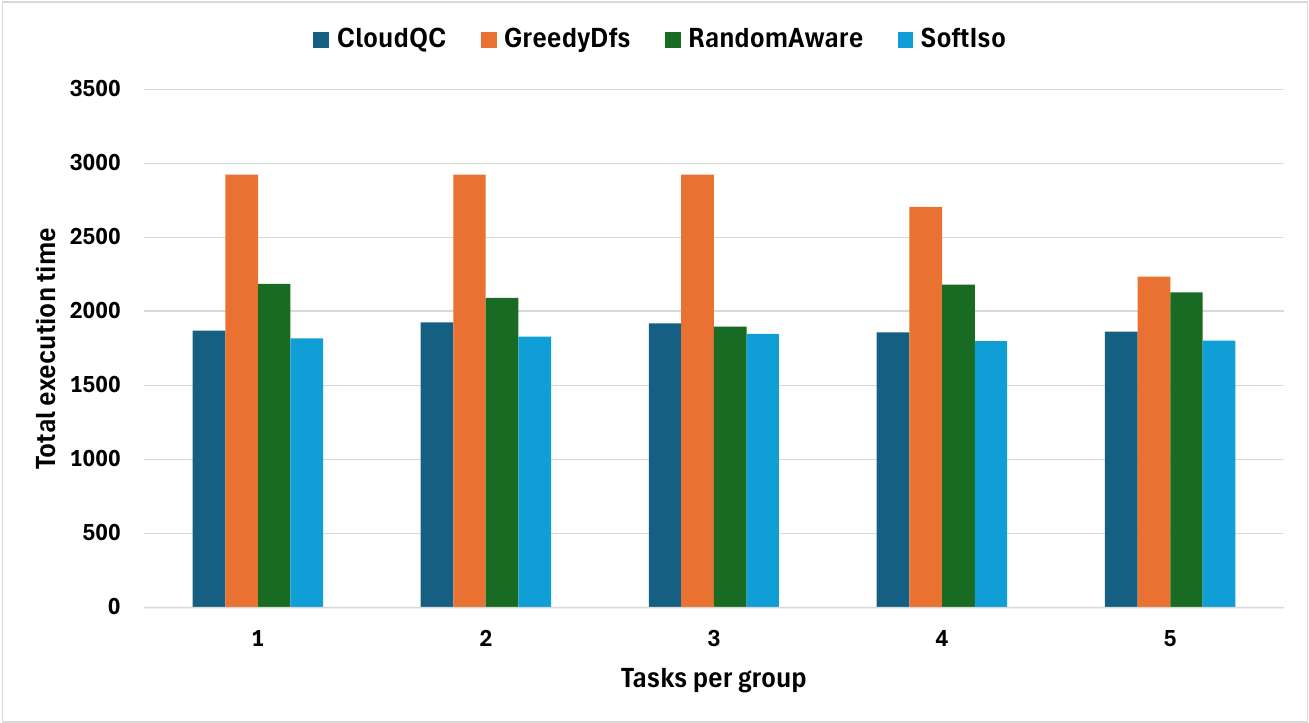}
                  \caption{}
         \label{fig:exec_task}
     \end{subfigure}
        \caption{Execution time(seconds) trends by (a) workload size for allocation, (b) number of nodes in the network and (c) distinct tasks per workflow}
        \label{fig:trend_exec}
\end{figure*}

\begin{figure*}[h]
     \centering
 \begin{subfigure}{0.33\linewidth}
         \centering
         \includegraphics[width=\linewidth, height=100pt]{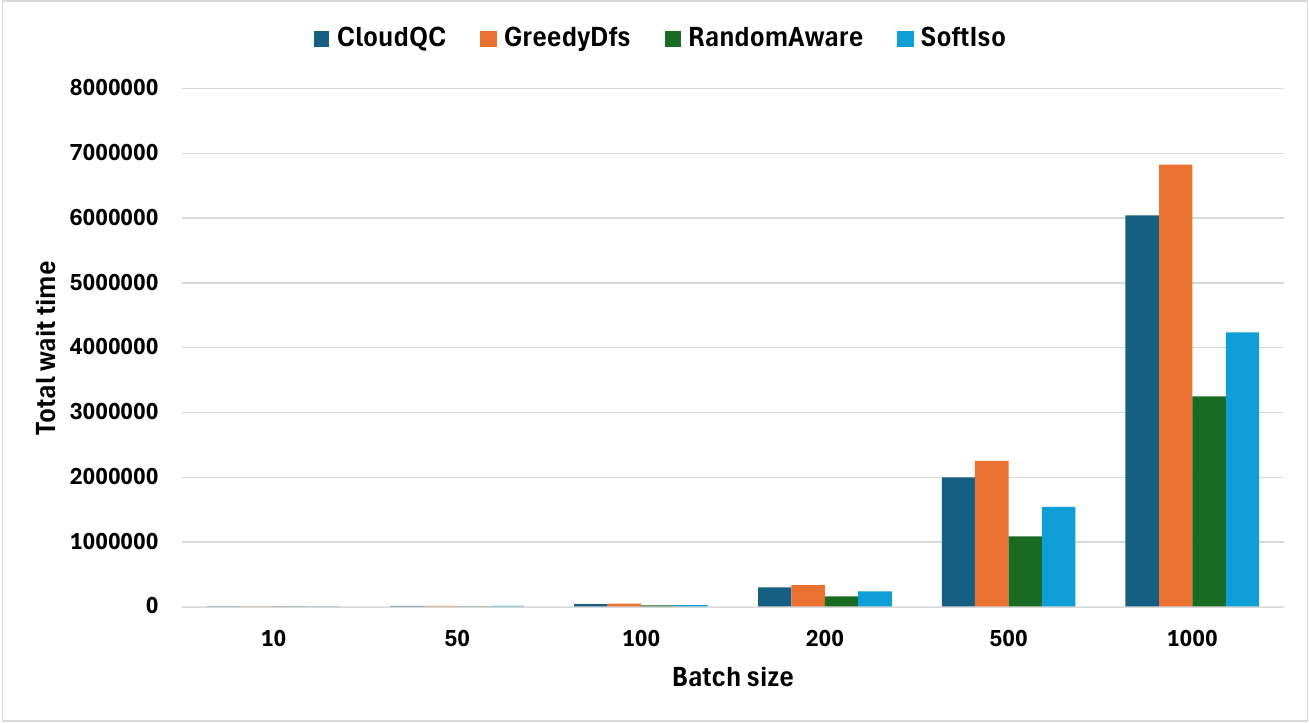}
                  \caption{}
         \label{fig:wait_batch}
     \end{subfigure}%
     \hfill
     \begin{subfigure}{0.33\linewidth}
         \centering
         \includegraphics[width=\linewidth, height=100pt]{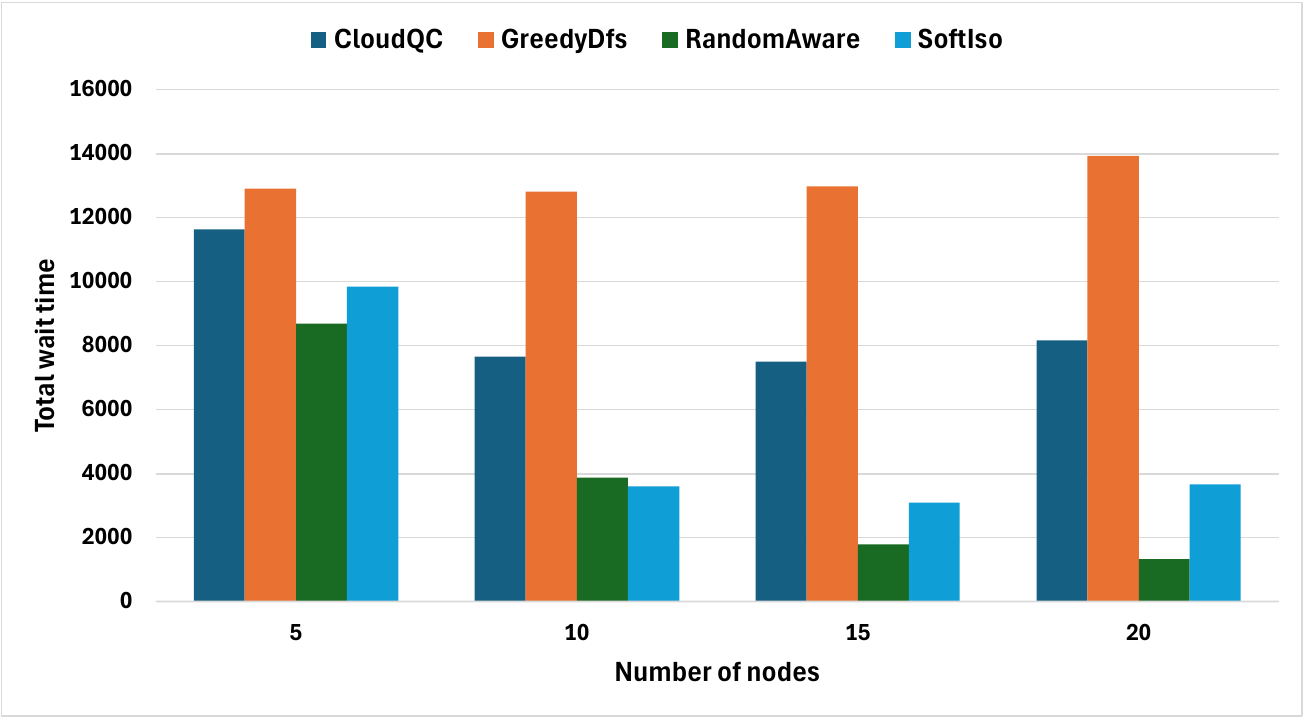}
                \caption{}
         \label{fig:wait_node}
     \end{subfigure}%
     \hfill
     \begin{subfigure}{0.33\linewidth}
         \centering
         \includegraphics[width=\linewidth, height=100pt]{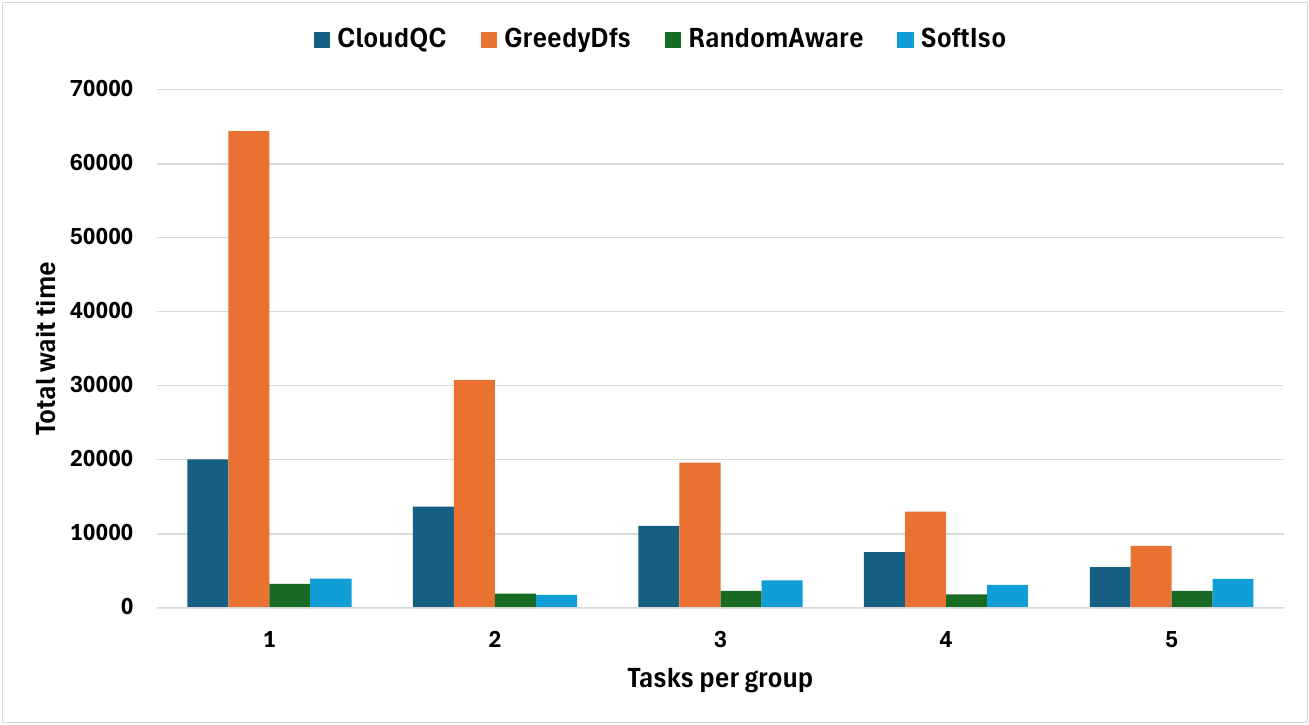}
                 \caption{}
         \label{fig:wait_task}
     \end{subfigure}
        \caption{Trends of wait time(seconds) by (a) workload size for allocation, (b) number of nodes in the network and (c) distinct tasks per workflow}
        \label{fig:trend_wait}
\end{figure*}

\subsection{Results and Discussion}
\label{sec: res}
We analyze the results observed from various experiments conducted for the allocation problem. This includes performance comparison of different algorithm, trend analysis under varying parameters, optimized cost evaluations, etc.

All the results reported in this section is an average of 100 experiments with similar parameters and different random seed for choices such  as, workflow selection, node topology, and other algorithm specific requirements.\\

\subsubsection{Trend Analysis}

We did experiments and analyze the performance trends over, the size of workload (batch size), the number of quantum machines available (nodes), and the number of tasks per workflow (group), as shown in sub figures (a), (b), (c), respectively of Figures \ref{fig:trend_comm}-\ref{fig:trend_wait}. 

The results here have (tasks per group, batch size) = (4, 50), (tasks per group, nodes) = (3, 5) \& (nodes, batch size) = (5, 50), for the analysis on the number of nodes, batch size and tasks per group, respectively. These values are chosen for a reasonable sized system and workflows which can be practically realized within the current infrastructure. 

As shown in Figures \ref{fig:com_batch}, \ref{fig:exec_batch} and \ref{fig:wait_batch}, the values of communication cost, execution time and wait times increases exponentially with the workload size. This is mainly because there are more tasks to execute and the cost of quantum simulation grows exponentially with problem size \cite{Georgescu2014QuantumSimulation}. On the other hand, fidelity generally increases linearly because of the linear relation between workload size and quantum program. A peak at batch size = 100 is observed because of the intricate choice of quantum nodes having less error rates for that experiment, see Figure. \ref{fig:fid_batch}.

With increase in the number of nodes, SoftIso and RandomAware algorithms have an increasing trend over different metrics. Whereas, CloudQC and GreedyDfs algorithms have either mixed or decreasing trends, see Figures \ref{fig:com_node}, \ref{fig:fid_node}, \ref{fig:exec_node}, \ref{fig:wait_node}. This shows that both SoftIso and RandomAware methods can be used effectively for larger environments.

Since the amount of links involved in communication increases with the number of tasks per workflow, there is an increasing trend seen in Figure. \ref{fig:com_task}. This linear increase supports well for the scalability of these solutions. But, the execution and wait times generally decrease with the number of tasks per group, see Figure. \ref{fig:exec_task}, \ref{fig:wait_task}. This is because of the simultaneous allocation of more tasks per decision cycle. Since a balanced cost function is used for Equation \ref{eq: costfunc}, fidelity and execution times had to compensate for this inherent increase in communication cost.

It is important for the allocator to evenly distribute the workload among all the available QPUs with a reasonable consideration of other cost metrics. An uniform distribution can help towards a scalable strategy that also avoids critical bottlenecks. As shown in Figure \ref{fig:dist_time}, the box plots represents the percentage of time invested by each QPU in executing the entire workload. Wider boxes represents an unfair distribution where some QPUs are overloaded than others. Plots having a smaller variance and a fairly central average line is best for our problem specification.

The impact of workload size on the variance of time shared by individual QPU is very insignificant. This is a good sign for all the algorithms in terms of scalability requirements. Though, the average share per QPU increases slightly with the workload size as there are more tasks to be executed. As shown in Figure \ref{fig:dt1_batch}, RandomAware performs best, with SofIso slightly better than CloudQC.

As the number of available nodes increase the load is distributed only to fewer QPUs while others remain idle, see Figure \ref{fig:dt1_node}. This indicates that there is no need to overwhelm the system with for too many QPUs for a reasonable allocation. When the number of nodes are set to 5 and 10, SoftIso performs better with an average load of 20\% per QPU and an evenly distributed workload. 

In Figure \ref{fig:dt1_task} we observe that most of the qpus remain idle for cases with fewer tasks per group. This is justified because workflows with less tasks require less networking as compared to workflows with more tasks. Since the average line tends towards 0, most of the qpus have less contribution when allocated using CloudQC. Whereas, for SoftIso and RandomAware algorithms the distributions are fair enough to most of the QPUs.\\

\subsubsection{Algorithm Comparisons}
\label{sec:res_sla}

\begin{table*}[h]
\centering
\caption{ Model performance based on average decision time and delivery(\%).
SP means small-program and LP means large-program with (tasks per
group, batch size) = (2, 10) and (4, 500), respectively. LR means less-resources and MR means more-resources with ($\varrho$, nodes) = (0.3, 10) and (0.9, 20), respectively.}

\begin{tabular}{|c|cccc||cccc|}

\toprule

\multirow{2}{*}{\textbf{Model}} & \multicolumn{4}{c||}{\textbf{Average completion (\%)}} & \multicolumn{4}{c|}{\textbf{Decision time (s)}}\\
\cmidrule(lr){2-5} \cmidrule(lr){6-9} 
& \textbf{SP-LR} & \textbf{SP-MR} & \textbf{LP-LR} & \textbf{LP-MR}  & \textbf{SP-LR} & \textbf{SP-MR}  & \textbf{LP-LR} & \textbf{LP-MR}\\

\midrule

CloudQC        & 100   & 100  & 32 & 100     
& 0.1281   & 0.2116  & 7.5288  & 9.8415 \\
GreedyDfs      & 100   & 100  & 30 & 100      
& 0.0206     & 0.0454  & 1.4315 & 0.9435  \\
RandomAware      & 100   & 100  & 10 & 99 
& 0.0337   & 0.0626 & 0.8018  & 2.0141 \\
SoftIso   & 100   & 100  & 91 & 100 
& 0.1933   & 0.3572 & 10.2132   & 74.1821\\

\bottomrule

\end{tabular}
\label{tab:performance}

\end{table*}

To understand and project a high level picture of the algorithmic performance, we created four different scenarios and did a stress test for all the algorithms. These cases include: 
\begin{itemize}
    \item SP-LR: Smaller system with smaller workflows, lesser workload and fewer nodes.
    \item SP-MR: Overwhelmed system with smaller workflows and batch sizes, but more quantum resources and a densely connected network.
    \item LP-LR: Overloaded system with much larger workload and complex workflows, but only a few quantum devices.
    \item LP-MR: Large system with enough resources and fully connected network capable of running large scale workloads.
\end{itemize}
Where, \textbf{SP} $\rightarrow$ small-program with (tasks per
group, batch size) = (2, 10),  \textbf{LP} $\rightarrow$ large-program with (tasks per
group, batch size) = (4, 500), \textbf{LR} $\rightarrow$ less-resources with ($\varrho$, nodes) = (0.3, 10), and \textbf{MR} $\rightarrow$ more-resources with ($\varrho$, nodes) = (0.9, 20). 

As reported in Table \ref{tab:performance}, all algorithms performs equally well in terms of completion percentage, except for the overloading case (LP-LR) where SoftIso performs exceptionally well. But, it should be noted that there is a trade-off in terms of cpu runtime for these techniques. Being a light weighted algorithm, RandomAware have the least decision time whereas, SoftIso takes the most amount of time due to the complexity of isomorphism problem. As compared to CloudQC, SoftIso has a reasonable trade-off with $\sim$3X improvement in completion percentage and around 50\% increase in cpu runtime. 

Even though the RandomAware algorithm has only 10\% completion rate for the overloading case, it is still the least time taking approach. If ran for more number of trials, this can achieve comparable or even better results than other algorithms within a reasonable time bound. GreedyDfs algorithm also performs better than CloudQC on these metrics, but as mentioned in the following section \ref{sec: cost_analysis} the trade-off over other critical cost metrics are not negligible.

\begin{figure}[h]
    \centering
    \includegraphics[width=\linewidth]{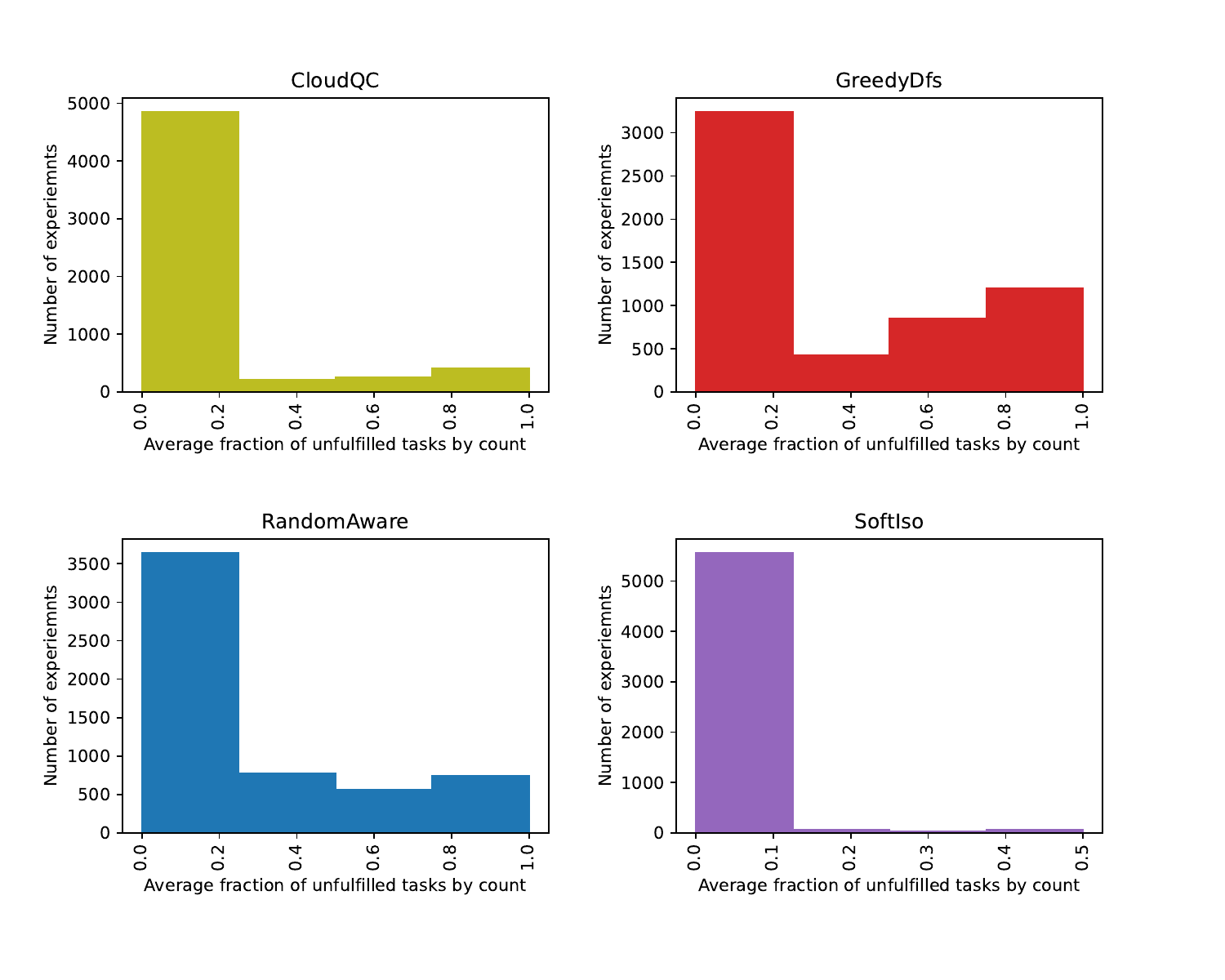}
    \caption{Histogram of unfulfilled tasks over all the experiments ran with different combinations of parameters.}
    \label{fig:sla_hist}
\end{figure}

If we see the distribution of task failures over all the 25K experiments, with every possible combinations of available hyperparameters (Figure \ref{fig:sla_hist}), it can be noted that SoftIso have reasonably higher completion rate. Where, other algorithms like GreedyDfs and RandomAware performs poorly with around 50\% - 30\% experiments having more than 25\% failure rate, SoftIso performs excellently with only around 2\% experiments with high failures rates. Even the CloudQC algorithm has almost 5X more experiments with more unfulfilled tasks, than the SoftIso algorithm.

\subsubsection{Costs Analysis}
\label{sec: cost_analysis}


After analyzing the execution times of the four algorithms, Figure \ref{fig:trend_exec}, it is clear that SoftIso performs better than all the other three techniques. It takes around 2-3\% less time than CloudQC. Being randomized, there are mixed results for RandomAware algorithm, but it does have improved results than SoftIso for few of the cases. On the contrary, RandomAware performs best in terms of wait time among all the tree algorithms. It is around 2X better than Cloud QC and 0.5X better than SoftIso, see Figure \ref{fig:trend_wait}.

Though there is not much difference in terms of fidelity among the four methods, Figure. \ref{fig:trend_fid}, SoftIso and RandomAware still have slightly higher values than CloudQC. But, when looked into the communication cost, see Figure \ref{fig:trend_comm}, SoftIso is a clear winner with $\sim$40-50\% improvement. RandomAware also has slightly better result than CloudQC with $\sim$5\% improvement in most of the cases. 

It is mainly because of the nature of our proposed methods to give equal priority to various cost functions that they performed better than CloudQC, which in principle focuses more on optimizing for connectivity. It is important to note that GreedyDfs performs worst in all the scenarios because it only solves for the constraints and doesn't consider the cost functions explicitly in the process.

\section{Conclusions and Future Work}
\label{sec: con}
In this paper, we proposed a practical solution to resource allocation problem for distributed quantum workflows using relevant system metrics into consideration. Our empirical results show that the proposed method, SoftIso, performs better than existing state-of-the-art methods, with average improvements of approximately 5\% in execution time, 30\% in communication overhead, 40\% in wait time and 2\% in fidelity, with a trade-off of around 50\% on cpu runtime. Even the alternative method, RandomAware, showed improved results in multiple scenarios.

Since development of quantum technology is an ongoing process, real life use cases for large scale problems can only be realized when proper infrastructure with quantum network is set up. Despite this fact, our work is still relevant to existing systems and can be easily extended to solutions with more metrics, workflow types, etc. This work can also be applied to environments with heterogeneous vendors and act as a supplementary module towards a complete solution for an end-to-end quantum cloud management system.

\section*{Acknowledgments}
This work is supported by the University of Melbourne and Maitri scholarships from the Department of Foreign Affairs and Trade, Government of Australia.

\bibliographystyle{ieeetr}
\bibliography{reference, bibl}

@book{papadimitriou1998combinatorial,
  title={Combinatorial optimization: algorithms and complexity},
  author={Papadimitriou, Christos H and Steiglitz, Kenneth},
  year={1998},
  publisher={Courier Corporation}
}

@inproceedings{ac1979some,
  title={Some complexity questions related to distributed computing},
  author={AC-C, YAO},
  booktitle={Proc. 11th Annual ACM Symposium on Theory of Computing, 1979},
  pages={209--213},
  year={1979}
}

@article{quetschlich2023mqtbench,
  title={{{MQT Bench}}: Benchmarking Software and Design Automation Tools for Quantum Computing},
  shorttitle = {{MQT Bench}},
  journal = {{Quantum}},
  author={Quetschlich, Nils and Burgholzer, Lukas and Wille, Robert},
  year={2023},
  note={{{MQT Bench}} is available at \url{https://www.cda.cit.tum.de/mqtbench/}},
}

@article{ibm-cal,
    title = {{IBM Callibration Data}},
    year = {2025},
    note = {Available at \url{https://quantum.cloud.ibm.com/docs/en/guides/qpu-information#calibration-data}}
}

@book{khang2024applications,
  title={Applications and principles of quantum computing},
  author={Khang, Alex},
  year={2024},
  publisher={IGI Global}
}

@article{Cordella2004AGraphs,
    title = {{A (sub)graph isomorphism algorithm for matching large graphs}},
    year = {2004},
    journal = {IEEE Transactions on Pattern Analysis and Machine Intelligence},
    author = {Cordella, Luigi P. and Foggia, Pasquale and Sansone, Carlo and Vento, Mario},
    number = {10},
    month = {10},
    pages = {1367--1372},
    volume = {26},
    url = {https://ieeexplore.ieee.org/abstract/document/1323804},
    doi = {10.1109/TPAMI.2004.75},
    issn = {01628828},
    pmid = {15641723}
}

@article{Qureshi2020ASystems,
    title = {{A comparative analysis of resource allocation schemes for real-time services in high-performance computing systems}},
    year = {2020},
    journal = {International Journal of Distributed Sensor Networks},
    author = {Qureshi, Muhammad Shuaib and Qureshi, Muhammad Bilal and Fayaz, Muhammad and Mashwani, Wali Khan and Belhaouari, Samir Brahim and Hassan, Saima and Shah, Asadullah},
    number = {8},
    month = {8},
    volume = {16},
    publisher = {SAGE Publications Ltd},
    url = {https://journals.sagepub.com/doi/full/10.1177/1550147720932750},
    doi = {10.1177/1550147720932750;WEBSITE:WEBSITE:SAGE;JOURNAL:JOURNAL:DSNA;WGROUP:STRING:PUBLICATION},
    issn = {15501477}
}

@article{Singh2017ASystems,
    title = {{A Survey and Comparative Study of Hard and Soft Real-Time Dynamic Resource Allocation Strategies for Multi-/Many-Core Systems}},
    year = {2017},
    journal = {ACM Computing Surveys (CSUR)},
    author = {Singh, Amit Kumar and Dziurzanski, Piotr and Mendis, Hashan Roshantha and Indrusiak, Leandro Soares},
    number = {2},
    month = {4},
    volume = {50},
    publisher = {ACMPUB27New York, NY, USA},
    url = {/doi/pdf/10.1145/3057267?download=true},
    doi = {10.1145/3057267},
    issn = {15577341}
}

@article{Hussain2013ASystems,
    title = {{A survey on resource allocation in high performance distributed computing systems}},
    year = {2013},
    journal = {Parallel Computing},
    author = {Hussain, Hameed and Malik, Saif Ur Rehman and Hameed, Abdul and Khan, Samee Ullah and Bickler, Gage and Min-Allah, Nasro and Qureshi, Muhammad Bilal and Zhang, Limin and Yongji, Wang and Ghani, Nasir and Kolodziej, Joanna and Zomaya, Albert Y. and Xu, Cheng Zhong and Balaji, Pavan and Vishnu, Abhinav and Pinel, Fredric and Pecero, Johnatan E. and Kliazovich, Dzmitry and Bouvry, Pascal and Li, Hongxiang and Wang, Lizhe and Chen, Dan and Rayes, Ammar},
    number = {11},
    month = {11},
    pages = {709--736},
    volume = {39},
    publisher = {North-Holland},
    url = {https://www.sciencedirect.com/science/article/abs/pii/S016781911300121X},
    doi = {10.1016/J.PARCO.2013.09.009},
    issn = {0167-8191}
}

@article{Professor2012AComputing,
    title = {{A Survey on Resource Allocation Strategies in                      Cloud Computing}},
    year = {2012},
    journal = {IJACSA) International Journal of Advanced Computer Science and Applications},
    author = {Professor, DrPadmavathiGanapathi},
    number = {6},
    volume = {3},
    url = {www.ijacsa.thesai.org}
}

@article{Krauter2002AComputing,
    title = {{A taxonomy and survey of grid resource management systems for distributed computing}},
    year = {2002},
    journal = {Software - Practice and Experience},
    author = {Krauter, Klaus and Buyya, Rajkumar and Maheswaran, Muthucumaru},
    number = {2},
    month = {2},
    pages = {135--164},
    volume = {32},
    publisher = {John Wiley {\&} Sons, Ltd},
    url = {/doi/pdf/10.1002/spe.432 https://onlinelibrary.wiley.com/doi/abs/10.1002/spe.432 https://onlinelibrary.wiley.com/doi/10.1002/spe.432},
    doi = {10.1002/SPE.432;PAGE:STRING:ARTICLE/CHAPTER},
    issn = {00380644}
}

@article{Ravi2021AdaptiveCloud,
    title = {{Adaptive job and resource management for the growing quantum cloud}},
    year = {2021},
    journal = {Proceedings - 2021 IEEE International Conference on Quantum Computing and Engineering, QCE 2021},
    author = {Ravi, Gokul Subramanian and Smith, Kaitlin N. and Murali, Prakash and Chong, Frederic T.},
    pages = {301--312},
    publisher = {Institute of Electrical and Electronics Engineers Inc.},
    isbn = {9781665416917},
    doi = {10.1109/QCE52317.2021.00047},
    arxivId = {2203.13260}
}

@article{Luo2025AdaptiveLearning,
    title = {{Adaptive Job Scheduling in Quantum Clouds Using Reinforcement Learning}},
    year = {2025},
    journal = {Proceedings of ACM Conference (Conference'17)},
    author = {Luo, Waylon and Zhao, Jiapeng and San Jose, Cisco and Zhan, Tong and Guan, Qiang},
    month = {6},
    volume = {1},
    url = {https://arxiv.org/abs/2506.10889v1},
    arxivId = {2506.10889}
}

@article{McKay2023BenchmarkingScale,
    title = {{Benchmarking Quantum Processor Performance at Scale}},
    year = {2023},
    author = {McKay, David C. and Hincks, Ian and Pritchett, Emily J. and Carroll, Malcolm and Govia, Luke C. G. and Merkel, Seth T.},
    month = {11},
    url = {https://arxiv.org/pdf/2311.05933},
    arxivId = {2311.05933}
}

@article{Corcoles2020ChallengesSystems,
    title = {{Challenges and Opportunities of Near-Term Quantum Computing Systems}},
    year = {2020},
    journal = {Proceedings of the IEEE},
    author = {Corcoles, Antonio D. and Kandala, Abhinav and Javadi-Abhari, Ali and McClure, Douglas T. and Cross, Andrew W. and Temme, Kristan and Nation, Paul D. and Steffen, Matthias and Gambetta, Jay M.},
    number = {8},
    month = {8},
    pages = {1338--1352},
    volume = {108},
    publisher = {Institute of Electrical and Electronics Engineers Inc.},
    url = {https://ieeexplore.ieee.org/abstract/document/8936946},
    doi = {10.1109/JPROC.2019.2954005},
    issn = {15582256},
    arxivId = {1910.02894}
}

@article{Magesan2012CharacterizingBenchmarking,
    title = {{Characterizing quantum gates via randomized benchmarking}},
    year = {2012},
    journal = {Physical Review A},
    author = {Magesan, Easwar and Gambetta, Jay M. and Emerson, Joseph},
    number = {4},
    month = {4},
    pages = {042311},
    volume = {85},
    publisher = {American Physical Society},
    url = {https://journals.aps.org/pra/abstract/10.1103/PhysRevA.85.042311},
    doi = {10.1103/PhysRevA.85.042311},
    issn = {10502947},
    arxivId = {1109.6887}
}

@article{Piveteau2024CircuitCommunication,
    title = {{Circuit Knitting With Classical Communication}},
    year = {2024},
    journal = {IEEE Transactions on Information Theory},
    author = {Piveteau, Christophe and Sutter, David},
    number = {4},
    month = {4},
    pages = {2734--2745},
    volume = {70},
    publisher = {Institute of Electrical and Electronics Engineers Inc.},
    doi = {10.1109/TIT.2023.3310797},
    issn = {15579654},
    arxivId = {2205.00016}
}

@article{Phillipson2023ClassificationComputing,
    title = {{Classification of Hybrid Quantum-Classical Computing}},
    year = {2023},
    journal = {Lecture Notes in Computer Science (including subseries Lecture Notes in Artificial Intelligence and Lecture Notes in Bioinformatics)},
    author = {Phillipson, Frank and Neumann, Niels and Wezeman, Robert},
    pages = {18--33},
    volume = {14077 LNCS},
    publisher = {Springer Science and Business Media Deutschland GmbH},
    url = {https://link.springer.com/chapter/10.1007/978-3-031-36030-5_2},
    isbn = {9783031360299},
    doi = {10.1007/978-3-031-36030-5{\_}2/TABLES/2},
    issn = {16113349},
    arxivId = {2210.15314}
}

@article{Zhou2025CloudQC:Computing,
    title = {{CloudQC: A Network-aware Framework for Multi-tenant Distributed Quantum Computing}},
    year = {2025},
    author = {Zhou, Ruilin and Gan, Yuhang and Liu, Yi and Qian, Chen},
    month = {4},
    url = {https://arxiv.org/pdf/2504.20389},
    arxivId = {2504.20389}
}

@article{CarreraVazquez2024CombiningCommunication,
    title = {{Combining quantum processors with real-time classical communication}},
    year = {2024},
    journal = {Nature},
    author = {Carrera Vazquez, Almudena and Tornow, Caroline and Rist{\`{e}}, Diego and Woerner, Stefan and Takita, Maika and Egger, Daniel J.},
    number = {8041},
    month = {12},
    pages = {75--79},
    volume = {636},
    publisher = {Nature Research},
    url = {https://www.nature.com/articles/s41586-024-08178-2},
    doi = {10.1038/S41586-024-08178-2;SUBJMETA},
    issn = {14764687},
    pmid = {39567696}
}

@article{Bova2021CommercialComputing,
    title = {{Commercial applications of quantum computing}},
    year = {2021},
    journal = {EPJ Quantum Technology 2021 8:1},
    author = {Bova, Francesco and Goldfarb, Avi and Melko, Roger G.},
    number = {1},
    month = {1},
    pages = {2-},
    volume = {8},
    publisher = {SpringerOpen},
    url = {https://link.springer.com/article/10.1140/epjqt/s40507-021-00091-1},
    doi = {10.1140/EPJQT/S40507-021-00091-1},
    issn = {2196-0763}
}

@article{Shi2020ConcurrentDesigns,
    title = {{Concurrent Entanglement Routing for Quantum Networks: Model and Designs}},
    year = {2020},
    journal = {SIGCOMM 2020 - Proceedings of the 2020 Annual Conference of the ACM Special Interest Group on Data Communication on the Applications, Technologies, Architectures, and Protocols for Computer Communication},
    author = {Shi, Shouqian and Qian, Chen},
    month = {7},
    pages = {62--75},
    publisher = {Association for Computing Machinery},
    url = {https://dl.acm.org/doi/pdf/10.1145/3387514.3405853},
    isbn = {9781450379557},
    doi = {10.1145/3387514.3405853;ISSUE:ISSUE:DOI}
}

@article{Dayarathna2016DataSurvey,
    title = {{Data center energy consumption modeling: A survey}},
    year = {2016},
    journal = {IEEE Communications Surveys and Tutorials},
    author = {Dayarathna, Miyuru and Wen, Yonggang and Fan, Rui},
    number = {1},
    month = {1},
    pages = {732--794},
    volume = {18},
    publisher = {Institute of Electrical and Electronics Engineers Inc.},
    url = {https://ieeexplore.ieee.org/abstract/document/7279063},
    doi = {10.1109/COMST.2015.2481183},
    issn = {1553877X}
}

@article{Buhrman2003DistributedComputing,
    title = {{Distributed Quantum Computing}},
    year = {2003},
    journal = {Lecture Notes in Computer Science (including subseries Lecture Notes in Artificial Intelligence and Lecture Notes in Bioinformatics)},
    author = {Buhrman, Harry and R{\"{o}}hrig, Hein},
    pages = {1--20},
    volume = {2747},
    publisher = {Springer, Berlin, Heidelberg},
    url = {https://link.springer.com/chapter/10.1007/978-3-540-45138-9_1},
    isbn = {978-3-540-45138-9},
    doi = {10.1007/978-3-540-45138-9{\_}1},
    issn = {1611-3349}
}

@article{Caleffi2024DistributedSurvey,
    title = {{Distributed quantum computing: A survey}},
    year = {2024},
    journal = {Computer Networks},
    author = {Caleffi, Marcello and Amoretti, Michele and Ferrari, Davide and Illiano, Jessica and Manzalini, Antonio and Cacciapuoti, Angela Sara},
    month = {12},
    pages = {110672},
    volume = {254},
    publisher = {Elsevier},
    url = {https://www.sciencedirect.com/science/article/pii/S1389128624005048},
    doi = {10.1016/J.COMNET.2024.110672},
    issn = {1389-1286},
    arxivId = {2212.10609}
}

@article{Boschero2025DistributedAndChallenges,
    title = {{Distributed Quantum Computing: Applications and Challenges}},
    year = {2025},
    journal = {Lecture Notes in Networks and Systems},
    author = {Boschero, Juan C. and Neumann, Niels M.P. and van der Schoot, Ward and Sijpesteijn, Thom and Wezeman, Robert},
    pages = {100--116},
    volume = {1423 LNNS},
    publisher = {Springer Science and Business Media Deutschland GmbH},
    url = {https://link.springer.com/chapter/10.1007/978-3-031-92602-0_6},
    isbn = {9783031926013},
    doi = {10.1007/978-3-031-92602-0{\_}6/FIGURES/1},
    issn = {23673389}
}

@article{Bisicchia2023DistributingByShots,
    title = {{Distributing Quantum Computations, by Shots}},
    year = {2023},
    journal = {Lecture Notes in Computer Science (including subseries Lecture Notes in Artificial Intelligence and Lecture Notes in Bioinformatics)},
    author = {Bisicchia, Giuseppe and Garc{\'{i}}a-Alonso, Jose and Murillo, Juan M. and Brogi, Antonio},
    pages = {363--377},
    volume = {14419 LNCS},
    publisher = {Springer Science and Business Media Deutschland GmbH},
    url = {https://link.springer.com/chapter/10.1007/978-3-031-48421-6_25},
    isbn = {9783031484209},
    doi = {10.1007/978-3-031-48421-6{\_}25/FIGURES/4},
    issn = {16113349}
}

@article{Nguyen2024DRLQ:Computing,
    title = {{DRLQ: A Deep Reinforcement Learning-based Task Placement for Quantum Cloud Computing}},
    year = {2024},
    journal = {IEEE International Conference on Cloud Computing, CLOUD},
    author = {Nguyen, Hoa T. and Usman, Muhammad and Buyya, Rajkumar},
    pages = {475--481},
    publisher = {IEEE Computer Society},
    isbn = {9798350368536},
    doi = {10.1109/CLOUD62652.2024.00060},
    issn = {21596190},
    arxivId = {2407.02748}
}

@article{Loke2022FromOverview,
    title = {{From Distributed Quantum Computing to Quantum Internet Computing: an Overview}},
    year = {2022},
    author = {Loke, Seng W.},
    month = {8},
    url = {https://arxiv.org/pdf/2208.10127},
    arxivId = {2208.10127}
}

@article{Nielson2010IntroductionInformation,
    title = {{Introduction - Quantum computation and quantum information}},
    year = {2010},
    author = {Nielson, M. A. and Chuang, Isaac L and Nielsen, Michael A and Chuang, Isaac L},
    pages = {700},
    url = {https://books.google.com/books/about/Quantum_Computation_and_Quantum_Informat.html?id=-s4DEy7o-a0C},
    isbn = {1139495488}
}

@article{Jaschke2023IsAdvantage,
    title = {{Is quantum computing green? An estimate for an energy-efficiency quantum advantage}},
    year = {2023},
    journal = {Quantum Science and Technology},
    author = {Jaschke, Daniel and Montangero, Simone},
    number = {2},
    month = {1},
    pages = {025001},
    volume = {8},
    publisher = {IOP Publishing},
    url = {https://iopscience.iop.org/article/10.1088/2058-9565/acae3e https://iopscience.iop.org/article/10.1088/2058-9565/acae3e/meta},
    doi = {10.1088/2058-9565/ACAE3E},
    issn = {2058-9565},
    arxivId = {2205.12092}
}

@article{Peterson1985OperatingConcepts,
    title = {{Operating system concepts}},
    year = {1985},
    author = {Peterson, James Lyle. and Silberschatz, Abraham.},
    pages = {625},
    publisher = {Addison-Wesley},
    isbn = {0201060892}
}

@article{Sane2025OptimizingApproach,
    title = {{Optimizing Resource Allocation in a Distributed Quantum Computing Cloud: A Game-Theoretic Approach}},
    year = {2025},
    author = {Sane, Bernard Ousmane and Hajdu{\v{s}}ek, Michal and Van Meter, Rodney},
    month = {4},
    url = {https://arxiv.org/abs/2504.18298v1},
    arxivId = {2504.18298}
}

@article{GiortamisOrchestratingQonductor,
    title = {{Orchestrating Quantum Cloud Environments with Qonductor}},
    author = {Giortamis, Emmanouil and Rom{\~{a}}o, Francisco and Tornow, Nathaniel and Lugovoy, Dmitry and Bhatotia, Pramod},
    arxivId = {2408.04312v1}
}

@article{Mantha2024Pilot-Quantum:Management,
    title = {{Pilot-Quantum: A Quantum-HPC Middleware for Resource, Workload and Task Management}},
    year = {2024},
    author = {Mantha, Pradeep and Kiwit, Florian J. and Saurabh, Nishant and Jha, Shantenu and Luckow, Andre},
    month = {12},
    url = {https://arxiv.org/abs/2412.18519v3},
    arxivId = {2412.18519}
}

@article{Nguyen2025QSimPy:Management,
    title = {{QSimPy: A learning-centric simulation framework for quantum cloud resource management}},
    year = {2025},
    journal = {Quantum Computing},
    author = {Nguyen, Hoa T. and Usman, Muhammad and Buyya, Rajkumar},
    month = {1},
    pages = {165--183},
    publisher = {Morgan Kaufmann},
    url = {https://www.sciencedirect.com/science/chapter/edited-volume/abs/pii/B978044329096100012X},
    isbn = {9780443290961},
    doi = {10.1016/B978-0-443-29096-1.00012-X},
    arxivId = {2405.01021}
}

@article{WackQualityComputers,
    title = {{Quality, speed, and scale: three key attributes to measure the performance of near-term quantum computers}},
    journal = {arxiv.orgA Wack, H Paik, A Javadi-Abhari, P Jurcevic, I Faro, JM Gambetta, BR JohnsonarXiv preprint arXiv:2110.14108, 2021•arxiv.org},
    author = {Wack, Andrew and Paik, Hanhee and Javadi-Abhari, Ali and Jurcevic, Petar and Faro, Ismael and Gambetta, Jay M and Johnson, Blake R},
    url = {https://arxiv.org/abs/2110.14108},
    arxivId = {2110.14108v2}
}

@article{Preskill2018QuantumBeyond,
    title = {{Quantum Computing in the NISQ era and beyond}},
    year = {2018},
    journal = {Quantum},
    author = {Preskill, John},
    month = {8},
    pages = {79},
    volume = {2},
    publisher = {Verein zur F{\"{o}}rderung des Open Access Publizierens in den Quantenwissenschaften},
    url = {https://quantum-journal.org/papers/q-2018-08-06-79/},
    doi = {10.22331/q-2018-08-06-79},
    issn = {2521327X},
    arxivId = {1801.00862v3}
}

@article{Qiao2022QuantumChallenges,
    title = {{Quantum Data Networking for Distributed Quantum Computing: Opportunities and Challenges}},
    year = {2022},
    journal = {INFOCOM WKSHPS 2022 - IEEE Conference on Computer Communications Workshops},
    author = {Qiao, Chunming and Zhao, Yangming and Zhao, Gongming and Xu, Hongli},
    publisher = {Institute of Electrical and Electronics Engineers Inc.},
    isbn = {9781665409261},
    doi = {10.1109/INFOCOMWKSHPS54753.2022.9798138}
}

@article{Childs2025QuantumConquer,
    title = {{Quantum Divide and Conquer}},
    year = {2025},
    journal = {ACM Transactions on Quantum Computing},
    author = {Childs, Andrew and Kothari, Robin and Kovacs-Deak, Matt and Sundaram, Aarthi and Wang, Daochen},
    number = {2},
    month = {4},
    volume = {6},
    publisher = {Association for Computing Machinery},
    url = {https://dl.acm.org/doi/10.1145/3723884},
    doi = {10.1145/3723884/ASSET/BFD644C9-F816-45AC-9434-437414A7531B/ASSETS/IMAGES/LARGE/TQC-2024-0001-EQN64.JPG},
    issn = {26436817},
    arxivId = {2210.06419}
}

@article{Cacciapuoti2020QuantumComputing,
    title = {{Quantum Internet: Networking Challenges in Distributed Quantum Computing}},
    year = {2020},
    journal = {IEEE Network},
    author = {Cacciapuoti, Angela Sara and Caleffi, Marcello and Tafuri, Francesco and Cataliotti, Francesco Saverio and Gherardini, Stefano and Bianchi, Giuseppe},
    number = {1},
    month = {1},
    pages = {137--143},
    volume = {34},
    publisher = {Institute of Electrical and Electronics Engineers Inc.},
    doi = {10.1109/MNET.001.1900092},
    issn = {1558156X},
    arxivId = {1810.08421}
}

@article{Georgescu2014QuantumSimulation,
    title = {{Quantum simulation}},
    year = {2014},
    journal = {Reviews of Modern Physics},
    author = {Georgescu, I. M. and Ashhab, S. and Nori, Franco},
    number = {1},
    month = {3},
    pages = {153},
    volume = {86},
    publisher = {American Physical Society},
    url = {https://journals.aps.org/rmp/abstract/10.1103/RevModPhys.86.153},
    doi = {10.1103/RevModPhys.86.153},
    issn = {15390756},
    arxivId = {1308.6253}
}

@article{Pompili2021RealizationQubits,
    title = {{Realization of a multinode quantum network of remote solid-state qubits}},
    year = {2021},
    journal = {Science},
    author = {Pompili, M. and Hermans, S. L.N. and Baier, S. and Beukers, H. K.C. and Humphreys, P. C. and Schouten, R. N. and Vermeulen, R. F.L. and Tiggelman, M. J. and dos Santos Martins, L. and Dirkse, B. and Wehner, S. and Hanson, R.},
    number = {6539},
    month = {4},
    pages = {259--264},
    volume = {372},
    publisher = {American Association for the Advancement of Science},
    url = {/doi/pdf/10.1126/science.abg1919},
    doi = {10.1126/SCIENCE.ABG1919;WEBSITE:WEBSITE:AAAS-SITE;JOURNAL:JOURNAL:SCIENCE;WGROUP:STRING:PUBLICATION},
    issn = {10959203},
    pmid = {33859028},
    arxivId = {2102.04471}
}

@article{Beukers2024Remote-EntanglementInterfaces,
    title = {{Remote-Entanglement Protocols for Stationary Qubits with Photonic Interfaces}},
    year = {2024},
    journal = {PRX Quantum},
    author = {Beukers, Hans K.C. and Pasini, Matteo and Choi, Hyeongrak and Englund, Dirk and Hanson, Ronald and Borregaard, Johannes},
    number = {1},
    month = {3},
    pages = {010202},
    volume = {5},
    publisher = {American Physical Society},
    url = {https://journals.aps.org/prxquantum/abstract/10.1103/PRXQuantum.5.010202},
    doi = {10.1103/PRXQuantum.5.010202},
    issn = {26913399}
}

@article{Endo2011ResourceChallenges,
    title = {{Resource allocation for distributed cloud: Concepts and research challenges}},
    year = {2011},
    journal = {IEEE Network},
    author = {Endo, Patricia Takako and De Almeida Palhares, Andre Vitor and Pereira, Nadilma Nunes and Goncalves, Glauco Estacio and Sadok, Djamel and Kelner, Judith and Melander, Bob and M{\aa}ngs, Jan Erik},
    number = {4},
    month = {7},
    pages = {42--46},
    volume = {25},
    url = {https://ieeexplore.ieee.org/document/5958007},
    doi = {10.1109/MNET.2011.5958007},
    issn = {08908044}
}

@article{Bahrani2024ResourceNetworks,
    title = {{Resource Management and Circuit Scheduling for Distributed Quantum Computing Interconnect Networks}},
    year = {2024},
    journal = {IEEE JOURNAL ON SELECTED AREAS IN COMMUNICATIONS},
    author = {Bahrani, Sima and Oliveira, Romerson D. and Parra-Ullauri, Juan Marcelo and Wang, Rui and Simeonidou, Dimitra},
    month = {9},
    volume = {XX},
    url = {https://arxiv.org/pdf/2409.12675},
    arxivId = {2409.12675}
}

@article{Jennings2014ResourceChallenges,
    title = {{Resource Management in Clouds: Survey and Research Challenges}},
    year = {2014},
    journal = {Journal of Network and Systems Management 2014 23:3},
    author = {Jennings, Brendan and Stadler, Rolf},
    number = {3},
    month = {3},
    pages = {567--619},
    volume = {23},
    publisher = {Springer},
    url = {https://link.springer.com/article/10.1007/s10922-014-9307-7},
    doi = {10.1007/S10922-014-9307-7},
    issn = {1573-7705}
}

@article{Goscinski1990ResourceSystems,
    title = {{Resource management in large distributed systems}},
    year = {1990},
    journal = {ACM SIGOPS Operating Systems Review},
    author = {Goscinski, Andrzej and Bearman, Mirion},
    number = {4},
    month = {9},
    pages = {7--25},
    volume = {24},
    publisher = {ACMPUB27New York, NY, USA},
    url = {https://dl.acm.org/doi/pdf/10.1145/94574.94575},
    doi = {10.1145/94574.94575},
    issn = {01635980}
}

@article{Barral2025ReviewComputing,
    title = {{Review of Distributed Quantum Computing: From single QPU to High Performance Quantum Computing}},
    year = {2025},
    journal = {Computer Science Review},
    author = {Barral, David and Cardama, F. Javier and D{\'{i}}az-Camacho, Guillermo and Fa{\'{i}}lde, Daniel and Llovo, Iago F. and Mussa-Juane, Mariamo and V{\'{a}}zquez-P{\'{e}}rez, Jorge and Villasuso, Juan and Pi{\~{n}}eiro, César and Costas, Natalia and Pichel, Juan C. and Pena, Tomás F. and G{\'{o}}mez, Andrés},
    month = {8},
    pages = {100747},
    volume = {57},
    publisher = {Elsevier},
    url = {https://www.sciencedirect.com/science/article/pii/S1574013725000231},
    doi = {10.1016/J.COSREV.2025.100747},
    issn = {1574-0137},
    arxivId = {2404.01265}
}

@article{Schmidt-Bleek2011TheIntervention,
    title = {{The Earth : Natural Resources and Human Intervention}},
    year = {2011},
    author = {Schmidt-Bleek, Friedrich.},
    pages = {247},
    publisher = {Haus Publishing},
    url = {https://books.google.com/books/about/The_Earth.html?id=BVcrDwAAQBAJ},
    isbn = {1906598592}
}

@article{Houbraken2014TheEnumeration,
    title = {{The Index-Based Subgraph Matching Algorithm with General Symmetries (ISMAGS): Exploiting Symmetry for Faster Subgraph Enumeration}},
    year = {2014},
    journal = {PLOS ONE},
    author = {Houbraken, Maarten and Demeyer, Sofie and Michoel, Tom and Audenaert, Pieter and Colle, Didier and Pickavet, Mario},
    number = {5},
    month = {5},
    pages = {e97896},
    volume = {9},
    publisher = {Public Library of Science},
    url = {https://journals.plos.org/plosone/article?id=10.1371/journal.pone.0097896},
    doi = {10.1371/JOURNAL.PONE.0097896},
    issn = {1932-6203},
    pmid = {24879305}
}


\end{document}